\documentclass{iopart}
\usepackage{iopams}
\usepackage{amssymb}
\usepackage{bm}
\usepackage{dsfont}
\usepackage{graphicx}
\usepackage{subfigure}
\usepackage{dcolumn}
\usepackage{braket}
\usepackage[usenames,dvipsnames,svgnames,table]{xcolor}

\newcommand{\ri}{\mathrm{i}}

\newcommand{\bo}{\hat{b}^{\phantom\dag}}
\newcommand{\ba}{\hat{b}^{\dag}}
\newcommand{\ao}{\hat{a}^{\phantom\dag}}
\renewcommand{\aa}{\hat{a}^{\dag}}

\newcommand{\no}{\hat{n}}
\newcommand{\Ho}{\hat{H}}

\newcommand{\Po}{\hat{P}}

\newcommand{\la}{\langle}
\newcommand{\ra}{\rangle}
\newcommand{\be}{\begin{equation}}
\newcommand{\ee}{\end{equation}}
\newcommand{\bes}{\begin{eqnarray}}
\newcommand{\ees}{\end{eqnarray}}

\newcommand{\dw}{\downarrow}
\newcommand{\up}{\uparrow}

\newcommand{\trace}{\mathrm{Tr}}
\newcommand{\text}[1]{\mathrm{#1}}

\begin{document}

\title[Quantum crystal growing]{Quantum crystal growing: Adiabatic preparation of a bosonic antiferromagnet in the presence of a parabolic inhomogeneity}
\author{S{\o}ren Gammelmark$^{1,2}$}
\author{Andr\'e Eckardt$^{1,3}$}
\address{$^1$ICFO-Institut de Ci\`encies Fot\`oniques, 
Av.\ Canal Ol\'impic s/n, E-08860 Castelldefels (Barcelona), Spain,}
\address{$^2$Lundbeck Foundation Theoretical Center for Quantum System Research,
Department of Physics and Astronomy, University of Aarhus, DK 8000 Aarhus
C, Denmark}
\address{$^3$Max-Planck-Institut f\"ur Physik komplexer Systeme, N\"othnitzer Stra\ss e 38, 
D-01187 Dresden, Germany}
\date{\today}

\begin{abstract}
We theoretically study the adiabatic preparation of an antiferromagnetic phase 
in a mixed Mott insulator of two bosonic atom species in a one-dimensional 
optical lattice. In such a system one can engineer a tunable parabolic 
inhomogeneity by controlling the difference of the trapping potentials felt by 
the two species. Using numerical simulations we predict that a finite 
parabolic potential can assist the adiabatic preparation of the 
antiferromagnet. The optimal strength of the parabolic inhomogeneity 
depends sensitively on the number imbalance between the two species. We also 
find that during the preparation finite size effects will play a crucial role
for a system of realistic size. The experiment that we propose can be 
realized, for example, using atomic mixtures of Rubidium 87 with Potassium 41 
or Ytterbium 168 with Ytterbium 174.
\end{abstract}
\maketitle

\section{Introduction}
The hardware of an (analog) quantum simulator \cite{Feynman82} is a controlled 
quantum system that is a clean and tunable realization of a many-body model 
system of interest (see also Refs.~\cite{HaukeEtAl12,BlochEtAl12}). 
In a quantum simulation such a quantum machine is used to 
experimentally measure dynamical or equilibrium properties of the model that 
are hard to obtain by using a classical machine. A typical protocol for 
studying equilibrium properties will start from a parameter regime of the 
model that is well understood theoretically and, thus, allows validating that 
a state close to thermal 
equilibrium can be prepared faithfully. In a next step, the system is then 
guided slowly into the parameter regime of interest. If it can be assumed that 
the dynamics during this parameter variation is close to adiabatic, the system 
will finally be in a state close to the target state, which is defined to be 
the thermal equilibrium for the new set of parameters characterized, e.g., by 
the same entropy and particle number as the initial state. 

However, typically a phase transition is expected to occur on the way between
the initial and the target regime. Crossing this transition is
potentially a source for an increased production of excitations as
described by the Kibble-Zurek mechanism in the case of a continuous phase 
transition (see Refs.~\cite{Dziarmaga10,PolkovnikovEtAl11} and
References therein). In order to keep defect creation at a minimal level, it
has been proposed to bring the system from one quantum phase to the other 
without passing a true phase transition by employing spatial inhomogeneity
\cite{DziarmagaEtAl99,DziarmagaRams10a,DziarmagaRams10b}. 
Namely, in an inhomogeneous system the transition from one phase to another 
can happen as a crossover at a spatial boundary of finite width. By parameter 
variation this boundary can be moved through the system at a finite speed, 
eventually bringing it from one phase to the other.  Such a strategy is 
similar to methods like growing crystals or pulling them out of the melt.
It has been investigated theoretically in the simple model system of a quantum 
spin-1/2 chain (with Ising or XY coupling) in an inhomogeneous transverse 
field; here the ferromagnet-to-paramagnet transition (a continuous phase 
transition in the uniform system) can be induced practically without defect 
creation, provided the phase boundary moves slow enough
\cite{DziarmagaRams10a,DziarmagaRams10b}. 

In this paper we investigate theoretically a problem of direct experimental 
relevance. Namely whether a parabolic inhomogeneity can be useful to assist 
the adiabatic preparation of an antiferromagnetic quantum phase in an 
experiment with ultra cold atoms 
\cite{BlochDalibardZwerger08,LewensteinSanperaAhufinger}.
The antiferromagnetic crystalline order shall be grown in space from the
center of the system outwards. To this end, we consider a two-species mixture of
ultra cold bosonic atoms in an optical lattice with strong on-site repulsion
(see, e.g., 
Refs.~\cite{AltmanEtAl03,DuanEtAl03,KuklovSvistunov03,HubenerEtAl09,Powell09,
Shrestha10,CapogrossoSansoneEtAl10,LiEtAl11} for the equilibirum properties of 
such a system).
The system can be described by a quantum XXZ-spin-1/2 model
\cite{AltmanEtAl03,DuanEtAl03,KuklovSvistunov03} and we are interested in the 
transition from an easy-plane ferromagnetic phase
in the spin $xy$ plane to an easy-axis antiferromagnetic phase in $z$
direction. We will concentrate on a one-dimensional system with the dynamics 
along the perpendicular directions frozen out by a strong transversal 
confinement.

The motivation for the present work is twofold. First of all, the quantum
antiferromagnetically ordered target state is known to be very fragile with
respect to thermal fluctuations, because it is stabilized by low-energy 
superexchange physics only \cite{CapogrossoSansoneEtAl10} (for 
antiferromagnetism in ultra cold atoms without superexchange cf. Refs.\ 
\cite{EckardtEtAl10,SimonEtAl11,StruckEtAl11,HaukeEtAl12b}). This makes its 
experimental realization challenging. It is, thus, desirable and of 
immediate relevance for current experimental studies to investigate how the
state can be prepared with a minimum of heating. A related problem of great 
importance is the preparation of the fermionic Heisenberg antiferromagnet 
being a prerequisite for mimicking the intriguing physics of high-temperature 
cuprate superconductors \cite{LeeEtAl06} with ultra cold atoms
\cite{HofstetterEtAl02}.

Another motivation lies in the fact that the system we are studying possesses
several interesting properties. It allows to experimentally control and 
(despite of the fact that the particles are always trapped) even switch off 
completely a parabolic inhomogeneity by tuning the relative trap
strength of the two bosonic species \cite{EckardtLewenstein10}. This enables 
the experimentalist to study the influence of (in)homogeneity in detail also 
in the laboratory. 

The system is also rich and generic enough to give rise to
effects that potentially disfavor the usage of inhomogeneity for the purpose
of an adiabatic state preparation. For example, mass flow can be a limiting
factor, especially if domains of insulating phases appear, acting as barriers
that hamper density redistribution, as recently discussed in the context of
the inhomogeneous bosonic Mott transition \cite{BernierEtAl11,NatuEtAl11}.
Another aspect is that the transition can change from a continuous
(second order) transition to a discontinuous (first order) transition (see
Ref.~\cite{HebertEtAl01} for the two-dimensional case).  
The continuous transition occurs without inhomogeneity, whereas the 
discontinuous transition is relevant for studying the transition with inhomogeneity.

For the related problem of a fermionic Heisenberg antiferromagnet adiabatic
protocols based on inhomogeneities that reduce the
discrete translational symmetry of the system, have been investigated recently
\cite{SoerensenEtAl10,LubaschEtAl11}.

In this paper, we study the influence of a static parabolic inhomogeneity, 
while the transition from one quantum phase to the other is induced by varying 
terms that themselves do not break translational symmetry. Similar scenarios 
have recently been investigated for temperature-driven phase transitions 
\cite{Zurek09,delCampoEtAl10,delCampoEtAl11} or for ramps not passing 
a phase transition \cite{HaqueZimmer12}. We are not considering a quench of 
the inhomogeneity itself, as it has been studied for example in 
Ref.~\cite{ColluraKarevski10}. We note in passing that the dynamics of a 
harmonically trapped bose-bose mixture in response to a sudden displacement of 
the trap has recently been investigated theoretically as a probe for different 
quantum phases in the system \cite{HuEtAl11}.

The numerical studies of the one-dimensional system that we present in this
paper indicate that a parabolic potential can indeed assist the adiabatic 
preparation of the antiferromagnetic target state. However, the optimal 
strength of the inhomogeneity depends in a sensitive way on the imbalance 
between the two bosonic species; the larger the imbalance the larger the 
optimal inhomogeneity. Therefore, the possibility to tune the inhomogeneity 
\cite{EckardtLewenstein10} should be an advantage for preparing the
antiferromagnetic order. We also observe that for realistic system sizes the 
time evolution during the parameter ramp into the antiferromagnetic regime is 
still governed by finite size effects that go beyond the local-density 
picture. Namely we find precursing antiferromagnetic correlations already in 
the ground state of the system outside the antiferromagnetic regime. These are 
contaminated with imperfections (like kinks) that originate from the 
inhomogeneity. The imperfect correlations are amplified when the system is 
ramped into the antiferromagnetic regime. It is difficult for the system to 
get rid of these imperfections, as it would be required for a perfectly 
adiabatic time evolution. So we find the best results for parameters giving 
rise to an initial state with a low degree of imperfections in the precursing 
antiferromagnetic order.

The paper is organized as follows. 
The system and the different models describing it are introduced in section 
\ref{sec:sys}.
The structure of the grand canonical ground state phase diagram is reviewed in section \ref{sec:pd}, 
with some details on how the phase diagram has been computed using the Bethe ansatz in appendix \ref{sec:Bethe}. The protocol for the 
preparation of the antiferromagnetic state is described in detail and 
motivated in section \ref{sec:prot}. The results of our numerical simulation 
of this protocol are finally presented in section \ref{sec:sim}, before we 
conclude in section \ref{sec:con}.

\section{\label{sec:sys}System and models}
We are considering a system of ultra cold atoms given by a mixture of two
bosonic species in one spatial dimension (1D) subjected to a steep optical
lattice potential. In recent experiments such mixtures have been
loaded into optical lattices, among them Potassium (K41) Rubidium (Rb87) 
mixtures  \cite{CataniEtAl08} and mixtures of different hyperfine (``spin'') 
states of Rb87 
\cite{MandelEtAl03b,LeeEtAl07,WeldEtAl09,GadwayEtAl10,MedleyEtAl11,
SoltanPanahiEtAl11}. Other candidates include mixtures of different 
Ytterbium-Isotopes \cite{FukuharaEtAl09, SugawaEtAl11} that offer a rich variety of scattering 
properties depending on the selection of isotopes \cite{KitagawaEtAl08}. In the following we consider a
two-species system characterized by all-repulsive interactions and with the 
intraspecies repulsion being strong compared to the interspecies repulsion. 
Such a system can be realized experimentally by using an Yb168-Yb174 mixture 
\cite{KitagawaEtAl08} or, alternatively, by taking a K41-Rb87 mixture
with the interspecies scattering length tuned small by means of a 
Feshbach-resonance \cite{ThalhammerEtAl09}. 

The 1D mixture of two bosonic species $s=a,b$ in an optical lattice is
described by the Bose-Hubbard Hamiltonian
\bes
\Ho_\text{BH} &=& \sum_{\ell=-\infty}^\infty 
\Big[  \sum_s \Big\{-J_s (\aa_{s\ell+1}\ao_{s\ell} + \text{h.c.})   
\nonumber\\&&
+\,(V_{s\ell}-\mu_s)\no_{s\ell} \Big\}
+ U_{ab}\no_{a\ell}\no_{b\ell}
\nonumber\\&&
+\,\frac{U_{aa}}{2}\no_{a\ell}(\no_{a\ell}-1)
+ \frac{U_{bb}}{2}\no_{b\ell}(\no_{b\ell}-1)\Big],
\ees
where $\ao_{s\ell}$ and $\no_{s\ell}$ are the bosonic annihilation and number 
operator for particles of species $s$ at lattice site $\ell$. Tunneling 
between neighboring sites is captured by the positive matrix elements $J_s$;  
the three positive Hubbard energies $U_{ab}$, $U_{aa}$, and $U_{bb}$ 
characterize the repulsive inter and intra species on-site interactions; and  
the particles are confined by the harmonic trapping potentials 
$V_{s\ell}=\frac{1}{2}\alpha_s\ell^2$. In the ground state the total numbers 
$N_a$ and $N_b$ of $a$ and $b$ particles are controlled by the chemical 
potentials $\mu_s$.

We are interested in the regime of strong repulsive interactions with the 
Hubbard energies $U_{s's}$ being positive and large compared to both the 
tunneling matrix elements $J_s$ and the chemical potentials $\mu_s$ such that 
double occupancy is strongly suppressed. Under these conditions we can 
effectively describe the system within the low-energy subspace $S$ defined by 
	\be
	S:\; \no_{a\ell}+\no_{b\ell}\le 1 \quad \forall \ell. 
	\ee
We employ degenerate-state perturbation theory \cite{Klein73} up to second 
order with respect to tunneling processes and obtain the effective Hamiltonian 
\bes\label{eq:Heff}
\Ho_\text{eff} &=& \sum_{\ell} \sum_s \Big[
(V_{s\ell}-\mu_s)\no_{s\ell} 
-J_s \Po_{S}(\aa_{s\ell+1}\ao_{s\ell} + \text{h.c.})\Po_{S}
\nonumber\\&&
-J\,\aa_{s\ell+1}\ao_{\bar{s}\ell+1}\aa_{\bar{s}\ell}\ao_{s\ell}
-\frac{U}{2}\no_{s\ell+1}\no_{\bar{s}\ell} \Big]
\ees
acting in $S$. Here the operator $\Po_{S}$ projects into the subspace $S$ and 
$\bar{s}$ denotes the species opposite to $s$. 
The first and second term originate from zeroth- and 
first-order perturbation theory, respectively. The new matrix elements $J$ and 
$U$ describe second-order superexchange processes.
While $J$ quantifies swaps between $a$ and $b$ particles on neighboring sites, 
$U$ characterizes an attractive nearest-neighbor interaction between $a$ and 
$b$ particles. These matrix elements stem from perturbative admixtures of 
Fock states with one site occupied by both an $a$ and a $b$ particle and read
\bes\label{eq:J}
J &=& 2\frac{J_aJ_b}{U_{ab}}
\\\label{eq:U}	
U &=& 2\frac{J_a^2+J_b^2}{U_{ab}}.
\ees

Furthermore, when deriving $\Ho_\text{eff}$ two additional approximations have 
been made for simplicity that both are well justified. The first one is 
that we neglected second-order terms involving virtual excitations 
(perturbative admixtures) with two particles of the same species on the same 
site. The amplitudes of such terms are proportional to 
$~J_{a(b)}^2/U_{aa(bb)}$ and are much smaller than the effective matrix 
elements (\ref{eq:J}) and (\ref{eq:U}), since we assume 
$U_{ab}\ll U_{aa},U_{bb}$.
The second simplification consists in neglecting the small potential energy 
differences between neighboring sites $(V_{s\ell+1}-V_{s\ell})= 
\alpha_{s}(\ell+1/2)$ that would appear together with $U_{ab}$ in the 
denominators of the second-order matrix elements (\ref{eq:J}) and 
(\ref{eq:U}). This approximation is well justified for typical slowly 
varying traps. 

For a Yb168-Yb174 mixture with scattering lengths 
$a_{168-168} = 13.33\ \mathrm{nm}$, $a_{174-168} = 0.13\ \mathrm{nm}$ and 
$a_{174-174} = 5.55\ \mathrm{nm}$ \cite{KitagawaEtAl08}, perpendicular 
lattice depth $V/E_r = 50$, transversal lattice depths $V_a/E_r = 16$, $V_b/E_r = 29$ and
lattice wave-length $\lambda = 532\ \mathrm{nm}$ leads to $J_a/U_{ab} \approx 0.12$ and $J_b/U_{ab} \approx 0.012$ corresponding approximately to Figure \ref{fig:full}(b) and a time-scale $\hslash/U \sim 26\ \mathrm{ms}$.

Concerning the level of approximation, the description of the bosonic system 
in terms of the effective Hamiltonian (\ref{eq:Heff}) is comparable to the 
$tJ$-model for spin-1/2 fermions on a lattice with strong on-site repulsion 
\cite{SpatekOles77,ZouAnderson88}. It describes the low energy physics of a
doped magnet by combing two elements; the superexchange coupling between the 
spin (or species) degree of freedom on neighboring occupied sites on the one 
hand and, on the other hand, the dynamics of the charge (or 
total density) degree of freedom due to the presence of holes (vacant
sites). For fermions the interplay between both is conjectured to give 
rise to intriguing physics like high-temperature superconductivity in the case 
of a square lattice \cite{LeeEtAl06}. The homogeneous version of the bosonic 
model (\ref{eq:Heff}) has been studied theoretically in 
Refs.~\cite{Boninsegni01,Boninsegni02,BoninsengiProkofev08} where, e.g., phase 
separation between hole-rich and hole-free regions is predicted on the square 
lattice. 

For slowly varying traps it is useful to introduce the local chemical 
potentials $\mu_{s\ell}\equiv\mu_s-V_{s\ell}$.
For sufficiently large  $\mu_a$ and $\mu_b$ (i.e.\ for a sufficiently large 
total particle number $N_a+N_b$), in an extended region $M$ in the trap center 
the local chemical potentials will be large enough to strongly suppress the 
existence of unoccupied sites.
(An estimate of the size of $M$ will be given at the 
end of this section). In this region the particles form a mixed Mott insulator 
with occupation $\la\no_a+\no_b\ra\simeq1$. The remaining degrees of 
freedom, namely which site is occupied by which species, can then effectively 
be described within the subspace $S'$ of unit filling,
\be
 S':\;\no_{a\ell}+\no_{\ell} = 1  \quad \forall \ell\in M .
\ee
In $S'$ and for $\ell\in M$ the Hamiltonian is again given by 
$\Ho_\text{eff}$, but the tunneling terms can now be dropped, 
giving 
\bes\label{eq:H}
\Ho &=& \sum_{\ell}
\Big[ -J(\aa_{a\ell+1}\ao_{b\ell+1}\aa_{b\ell}\ao_{a\ell}+\text{h.c.})
\nonumber\\&&
-\,\frac{U}{2}(\no_{b\ell+1}\no_{a\ell} +\no_{a\ell+1}\no_{b\ell})
+\frac{1}{2}(V_{\ell}-\mu)(\no_{a\ell}-\no_{b\ell})\Big].
\nonumber\\
\ees
We have also omitted the constant terms 
$\frac{1}{2}(\mu_{a\ell}+\mu_{b\ell})(\no_{a\ell}+\no_{b\ell})=\frac{1}{2}
(\mu_{a\ell}+\mu_{b\ell})$ and introduced 
the notation 
\be\label{eq:ih}
V_\ell\equiv V_{a\ell}-V_{b\ell}=\frac{1}{2}(\alpha_a-\alpha_b)\ell^2
\equiv\frac{1}{2}\alpha\ell^2
\ee
and
\be
\mu\equiv\mu_a-\mu_b.
\ee 
The effective Hamiltonian (\ref{eq:H}) will be the starting point for the 
remaining sections of this paper. 
  
The inhomogeneity $V_\ell$ appearing in $\Ho$ is characterized by the 
\emph{difference} of the trap frequencies $\alpha=\alpha_a-\alpha_b$.
This can be explained as a consequence of the constraint 
$\no_{a\ell}+\no_{b\ell}=1$; tunneling of an $a$ particle from 
$\ell$ to $\ell'$ has to be combined with the counterflow of a $b$ particle 
from $\ell'$ to $\ell$. In an experiment the degree of inhomogeneity $\alpha$ 
can be tuned continuously, simply by adjusting the trapping potentials of the 
two species with respect to each other. In particular, the accessible 
parameter space contains the \emph{homogeneous} model with $\alpha=0$ that is 
realized for equal traps $\alpha_1=\alpha_2$ \cite{EckardtLewenstein10} (as 
well as the regime of $\alpha<0$). The 
fact that this limit can be reached without the model description breaking 
down is a crucial ingredient of the experiment proposed here. 

Apart from the 
dimensionless inhomogeneity $\alpha/U$, the Hamiltonian $\Ho$ describing the 
simulator region $M$ of our system is characterized by two further 
dimensionless parameters, namely $J/U=(J_a/J_b+J_b/J_a)^{-1}$ and $\mu/U$. 
In an experiment these can be controlled independently by adjusting the ratio 
of tunneling strengths $J_a/J_b$ (controlled by the lattice depths for $a$ and 
$b$ particles) and the imbalance between $a$ and $b$ particles 
$(N_a-N_b)/(N_a+N_b)$.

It is both instructive and convenient to express the Hamiltonian (\ref{eq:H}) 
in terms of composite-particle \cite{LewensteinEtAl04} and spin 
\cite{KuklovSvistunov03} degrees of freedom. The former description is 
obtained by introducing composite particles that are hard-core bosons with 
annihilation operators  
\be
\bo_\ell=\aa_{b\ell}\ao_{a\ell}
\ee
and number operators $\no_\ell\equiv\ba_\ell\bo_\ell$,
such that $[\bo_\ell,\ba_{\ell'}]=\delta_{\ell,\ell'}(1-2\no_\ell)$ and
\be
	\bo_\ell\bo_\ell=\ba_\ell\ba_\ell=0  
\ee
in $S'$. With that $\no_\ell\equiv\ba_\ell\bo_\ell \le 1$. In $S'$
the composite-particle occupation numbers are equal to those of the 
$a$ particles, $\no_{\ell}=\no_{a\ell}(\no_{b\ell}+1)=\no_{a\ell}$,
whereas ``composite holes'' correspond to $b$ particles, $1-\no_\ell=\no_{b\ell}$. 
The Hamiltonian (\ref{eq:H}) can now be re-written like
\bes\label{eq:Hb}
\Ho &=& \sum_{\ell}
\Big[ -J(\ba_{\ell+1}\bo_{\ell}+\text{h.c.})
+U \Big(\no_{\ell+1}-\frac{1}{2}\Big)\Big(\no_{\ell}-\frac{1}{2}\Big)
\nonumber\\&&
+\,(V_{\ell}-\mu)\Big(\no_{\ell}-\frac{1}{2}\Big)\Big].
\ees
This Hamiltonian describes hard-core bosons in a tunable trapping potential 
$V_\ell$, with hopping matrix element $J$, \emph{repulsive} nearest 
neighbor interactions $U$, and chemical potential $\mu$. 

A spin-1/2 description is defined by identifying the species $s$ with an 
internal spin degree of freedom with spin $\up$ ($\dw$) for $a$ ($b$) 
particles. Introducing the vector of Pauli matrices ${\bm \sigma}_{s's}$ for 
this spin degree of freedom, we can define the spin operator at site $\ell$: 
	\be
	\hat{\bm S}_\ell 
	= \frac{1}{2}\sum_{s's}\aa_{s'\ell}{\bm \sigma}_{s's}\ao_{s\ell}
	\ee
with components
	\bes
	\hat{S}^x_{\ell}&=& \frac{1}{2}(\aa_{a\ell}\ao_{b\ell}+ \text{h.c.} )
	\\
	\hat{S}^y_{\ell}&=& \frac{1}{2}(\ri\aa_{b\ell}\ao_{a\ell}+ \text{h.c.})
	\\
	\hat{S}^z_{\ell}&=& \frac{1}{2}(\no_{a\ell}-\no_{b\ell}).
	\ees
In the subspace $S'$ these spin operators $\hat{\bm S}_\ell $ describe 
a spin-1/2 degree of freedom at every site. In terms of these degrees of 
freedom the Hamiltonian takes the form
	\be\label{eq:Hs}
	\Ho= \sum_\ell\Big[						
		-2J(\hat{S}^x_{\ell}\hat{S}^x_{\ell+1}
			+\hat{S}^y_{\ell}\hat{S}^y_{\ell+1})
		+ U 	\hat{S}^z_{\ell}\hat{S}^z_{\ell}	
		+(V_\ell-\mu)\hat{S}^z_{\ell}	\Big]
	\ee  
of an XXZ spin chain with ferromagnetic spin coupling $-2J\equiv 
J_x=J_y$ in $x$ and $y$ direction, antiferromagnetic Ising coupling 
$+U\equiv J_z$ in $z$ direction, and an inhomogeneous magnetic field 
$(V_\ell-\mu)\equiv h_\ell$ in $z$ direction. 

In the following we will focus on the central Mott insulator region $M$ 
described by the Hamiltonian $\Ho$. It serves as a simulator for the 
dynamics described by the model Hamiltonian 
$\Ho$ with tunable parabolic inhomogeneity. Let us briefly estimate the size 
of the Mott insulator region in the zero-temperature equilibrium state. In the limit of zero 
tunneling both species fill up the 
trap such that sites $\ell$ with $|\ell|<(N_a+N_b)/2\equiv \ell_0$ are 
occupied. If $\mu_{s^*}$ is the larger of the two chemical potentials $\mu_a$ 
and $\mu_b$, and $\alpha_{s^*}$ denotes the corresponding trap frequency, one 
has $\frac{1}{2}\alpha_{s^*}\ell_0^2=\mu_{s^*}$. In order to suppress doubly 
occupied sites near the trap center besides $J_s\ll U_{ab}$ also
$\mu_{s^*}<U_{ab}$ is required. The latter implies that
$2\ell_0<\sqrt{8 U_{ab}/\alpha_{s^*}}$. 

When finite tunneling is included, the edge 
of the occupied region will soften. With increasing $|\ell|$, near 
$|\ell|=\ell_0$ the occupation $\la\no_a+\no_b\ra$ will drop from 1 to 0 
within a crossover region of width $\Delta\ell$. The width $\Delta\ell$ is 
such that the increase of the trapping potential in the crossover region is of 
the order of the tunneling matrix element. More precisely,  $\partial_\ell 
V_{s\ell}|_{\ell=\ell_0}\Delta\ell_s=\alpha_\ell\ell_0\Delta\ell_s\sim J_s$
gives $\Delta\ell_s\sim J_s/(\alpha_s\ell_0)$ and one has either 
$\Delta\ell=\max(\Delta\ell_a,\Delta\ell_b)$ if both species can be found
near the edge $|\ell|\approx\ell_0$ (that is $\mu_a\approx\mu_b$) or
$\Delta\ell=\Delta_{s^*}$ if basically only $s^*$ particles occupy the edge 
region (i.e.\ $\mu_{\bar{s}^*}\ll\mu_{s^*}$). In the former (more restrictive) 
case the crossover region $\Delta\ell$ will be small compared to $\ell_0$ 
as long as $J_s\ll\alpha_s\ell_0^2\approx2\mu_{s^*}<2U_{ab}$, which has been 
required already. Therefore, the ``simulator region'' $M$ has an extent 
of $L\sim2\ell_0-2\Delta\ell$ that will be of the order of $\ell_0$ and it
can host an extensive fraction of the particles in the system.

\section{\label{sec:pd} Ground state phase diagram of the homogeneous system}

\begin{figure}
 \centering
   \includegraphics[width=1\columnwidth]{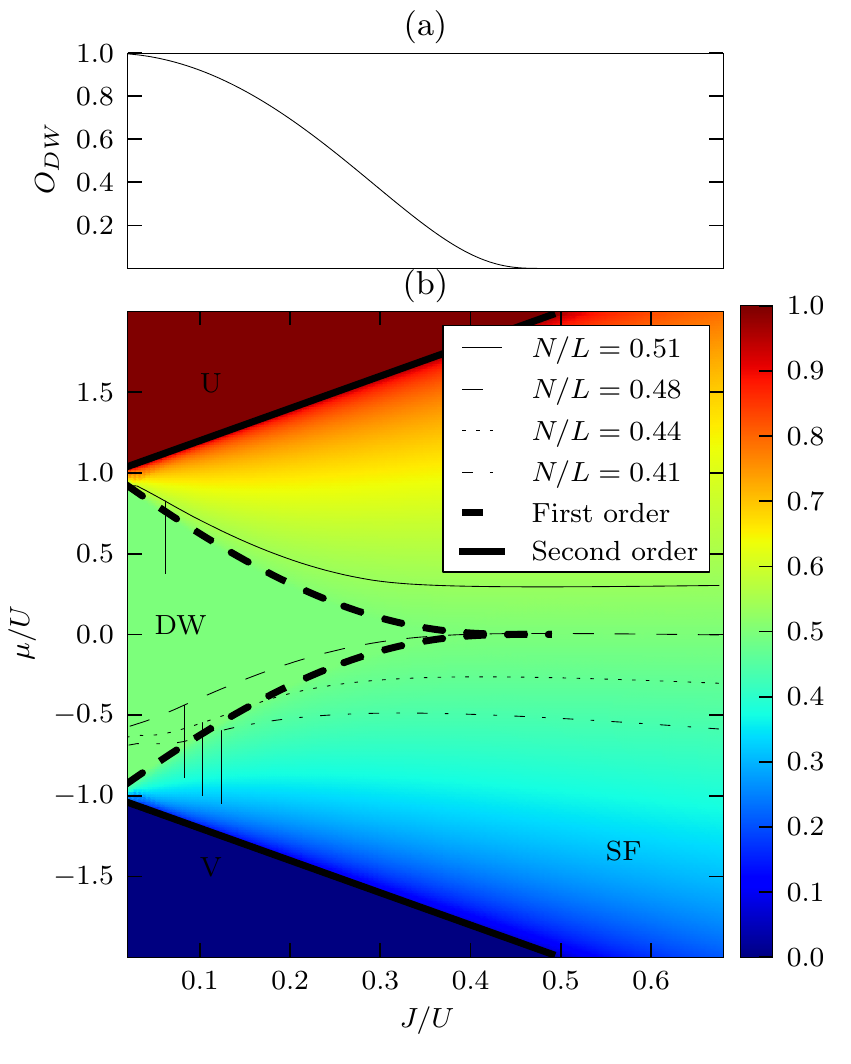}
    \caption{\label{fig:pd} 
Ground-state phase diagram of hard-core bosons in a one-dimensional 
lattice with nearest-neighbor repulsion $U$ described by the
Hamiltonian (\ref{eq:Hb}) with $V_\ell=0$. Subfigure (b) shows the parameter 
plane given by the dimensionless chemical potential $\mu/U$ and tunneling 
parameter $J/U$. 
Thick solid (dashed) lines indicate continuous (discontinuous) phase
transitions. The system 
possesses a compressible superfluid phase (SF) and three incompressible 
phases, the vacuum at zero filling (V), the particle-hole reflected vacuum 
at unit filling (U), and a density-wave insulator at half filling (DW). The 
color code describes the filling $n$ and clearly indicates the incompressible 
regions where $\partial n/\partial\mu=0$. 
The DW insulator is characterized by a
long-ranged staggered order in the site occupation, measured by the order 
parameter $O_\text{DW}$ defined in Eq.~(\ref{eq:ODW}). $O_\text{DW}$ is 
non-zero only in the DW phase, where it depends on $J/U$, as plotted in 
subfigure (a), but not on $\mu/U$. The sudden jump of $O_\text{DW}$ to zero 
when entering the SF phase renders the DW-to-SF transition discontinuous 
(though the filling $n$ varies continuously when passing the 
transition). The only exception occurs at the tip of the lobe-shaped DW domain 
(located at $\mu/U=0$ and $J/U=1/2$) where the transition is continuous.
The different thin curves plotted in (b) indicate the chemical potential 
$\mu/U$ of finite systems in the local density approximation with 
$N=31,29,27,25$ particles on $L=61$ sites 
subjected to a trap of strength $\alpha/U=10^{-3}$. The vertical ``system 
lines'' attached to these curves indicate how much the chemical potential
drops between the center and the edge of the system according to the local 
density approximation.} 
\end{figure}

In the central Mott region $M$ the system is described by the Hamiltonian 
$\Ho$ that we expressed in different representations. In the following we will 
stick to the language of the hard-core boson model (\ref{eq:Hb}), unless we 
explicitly mention the two-species [Eq.~(\ref{eq:H})] or spin 
[Eq.~(\ref{eq:Hs})] description. For a homogeneous system with $V_\ell=0$, the 
ground state of this model is characterized by two dimensionless parameters, 
the scaled chemical potential $\mu/U$ and the scaled tunneling matrix element 
$J/U$, with the nearest-neighbor repulsion $U$ serving as the unit of energy. 
We have computed the phase diagram of the homogeneous system in the 
$\mu/U$-$J/U$ plane by employing the Bethe-Ansatz solution developed in Refs.\ 
\cite{YangYang66a,YangYang66b,YangYang66c}.
Details of this solution are given in appendix \ref{sec:Bethe}.
In Fig.~\ref{fig:pd}(b) we plot the zero-temperature phase diagram and
the boson filling $n=\frac{1}{L}\sum_{\ell\in 
M}\la\no_i\ra$ with $L=\sum_{\ell\in 
M}1$ denoting the number of lattice sites in 
$M$.

The phase diagram shown in Fig.~\ref{fig:pd}(b) possesses the following 
structure.
First of all, it reflects the particle-hole symmetry of the homogeneous 
hard-core boson model (\ref{eq:Hb}); replacing $\bo_\ell\to\ba_\ell$ [implying
$(\no_\ell-1/2)\to (1/2-\no_\ell)$] and $\mu\to-\mu$ leaves the Hamiltonian 
unchanged, such that $\mu\to-\mu$ simply interchanges the role of particles 
and holes. 

The energy of a single particle with respect to the vacuum energy is given by 
$-\mu-U-2J$ with the kinetic energy reduction $-2J$ stemming from 
delocalization. Therefore, below a 
chemical potential of $\mu_v/U=-2J/U-1$ the system is in the vacuum (V) 
state $|v\ra$ with no particles present. Accordingly, for chemical potentials 
larger than $\mu_u/U=-\mu_v/U=2J/U+1$ the ground state is the particle-hole 
reflected vacuum, that is the incompressible insulating state 
$|u\ra=\prod_\ell\ba_\ell|v\ra$ at unit filling (U) with exactly one hard-core 
particle at every site. 

Starting from the vacuum state and increasing the 
chemical potential $\mu/U$, for non-zero tunneling $J/U$ the particle number 
starts to grow in a continuous fashion once the critical parameter $\mu_v/U$ 
is passed. Here the system enters a superfluid (SF) phase in a second-order 
phase transition.  This phase is characterized by a finite compressibility 
$\partial_\mu n\ne0$, a homogeneous density distribution 
$n_\ell=\la\no_\ell\ra=n$, and quasi-long-range off-diagonal order, i.e.\ 
the correlation function $\la\ba_\ell\bo_{\ell'}\ra$ decays algebraically 
for large $|\ell-\ell'|$.

For both chemical potential $\mu/U$ and tunneling $J/U$ small, another  
incompressible phase is found at half filling, a density wave (DW) Mott 
insulator [see Fig.~\ref{fig:pd}(b)]. This phase can be understood by starting 
from the limit of zero tunneling $J/U=0$. Here for $-1<\mu/U<1$ a DW state with one particle 
on every other site is favored as ground state. This state is two-fold 
degenerate and breaks the translational symmetry of the lattice.  
It possesses an energy gap $\Delta=\min(\Delta_p,\Delta_h)$ where 
$\Delta_p=U-\mu$ and $\Delta_h=\mu-U$ are the energy costs for adding a 
particle or adding a hole (removing a particle) somewhere in the system, 
respectively. [Particle number conserving particle-hole excitations
(created e.g. if one particle tunnels to a neighboring site) come with the 
larger energy cost $\tilde{\Delta}=\Delta_p+\Delta_h=2U$.] Since the 
Hamiltonian (\ref{eq:Hb}) conserves the total particle number the gap $\Delta$
protects a state at half filling from competing states with different particle 
numbers also for finite tunneling $J/U$, roughly as long as $\Delta$ is 
larger than the delocalization energy $2J$. The rough estimate $\Delta=2J$ for 
the phase boundary (corresponding to first order perturbation theory with 
respect to tunneling) explains the lobe shape of the DW insulator domain in 
the phase diagram  and (accidentally) even gives the correct critical 
tunneling strength $(J/U)_c=1/2$ at the tip of the lobe. Actually, this value 
of 1/2 is fixed by symmetry, namely as the Heisenberg point of the model 
in spin representation [Eq.~(\ref{eq:Hs})] where $|J_x|=|J_y|=|J_z|$.\footnote{
Re-defining $\hat{S}^z_\ell\to-\hat{S}^z_\ell$ on every 
other site also the Ising coupling in $z$-direction becomes ferromagnetic. Now 
the DW order corresponds to (long-range) ferromagnetic order in $z$ direction 
and the SF phase to (quasi-long-range) ferromagentic order in the $xy$ plane. 
At the Heisenberg point the system is known to possess spin-isotropic 
quasi-long-range ferromagnetic order. Now increasing/decreasing the 
$z$-coupling relative to the $xy$-coupling slightly, an easy axis/plane is 
created that immediately attracts (at least part of) the ferromagnetic 
correlations, guaranteeing DW/SF order. (This argument does not exclude an 
intermediate supersolid phase with both orders present, however such a phase 
is not found within the Bethe ansatz solution.)}
 
Within the DW domain the particle number and with that the whole structure of 
the ground state does not depend on the chemical potential $\mu/U$. Thus, 
(within this domain) also the DW order is a function of $J/U$ only. 
It can be quantified in terms of the long-ranged density-density correlations
by using the order parameter 
\be\label{eq:ODW}
O_\text{DW}=\lim_{|\ell-\ell'|\to\infty}
	\la(2\no_\ell-1)(2\no_{\ell'}-1)\ra(-1)^{\ell-\ell'}
\ee
that assumes values between 0 and 1. An analytical expression 
\cite{Sutherland2004Beautiful}\footnote{There is a misprint in \cite[p.\ 
186]{Sutherland2004Beautiful}. In the formula immediately preceding Eq.~(245) 
$\sigma^2$ should be $\sigma^4$. } is given by
\be\label{eq:ODWexp}
O_\text{DW} = \prod_{n=1}^\infty \tanh^4( n \lambda ), \quad \cosh(\lambda) = U/2J
\ee
and plotted in Fig.~\ref{fig:pd}(a). The fact that the order 
parameter $O_\text{DW}$ depends on $J/U$ only implies immediately that 
$O_\text{DW}$ drops in a discontinuous fashion from a finite value to zero
when the phase boundary of the DW domain is passed. This makes the transition
from DW to SF first order almost everywhere. As the only exception, the DW-SF
transition becomes continuous (second order) through the tip of the 
lobe; this describes the transition at fixed particle number (half filling) 
driven by tunneling.
Note that unlike the case of a two-dimensional square lattice with true
long-range order in the SF phase \cite{HebertEtAl01}, where also the particle 
number varies in a discontinuous fashion at the DW-SF transition, 
in one dimension the filling $\la\no\ra$ continuously departs from $1/2$ when
entering the SF phase. 

All in all, at $T=0$ the model possesses four different phases. A homogeneous 
SF phase and three distinct insulating phases, the vacuum (V) with $n_\ell=0$, 
the particle-hole reflected vacuum with unity filling (U) $n_\ell=1$, and the 
DW Mott insulator at half filling $n=1/2$ with alternating site occupations. 
In the original two-species picture (\ref{eq:H}), the $n=1$ and $n=0$ insulators correspond 
to a Mott insulator state consisting solely of $a$ or $b$ particles, 
respectively; the DW insulator is a Mott insulator with a staggered 
configuration of $a$ and $b$ particles, and the SF phase is a 
counterflow superfluid where a superflow of $a$ particles is accompanied by 
the corresponding back flow of $b$ particles such that $\la\no_a+\no_b\ra=1$. 
Finally, in the spin language (\ref{eq:Hs}) the $n=0$ and $n=1$ insulators 
correspond to the fully $z$-polarized state, the DW state is a phase with 
antiferromagnetic long-range order in the $z$-components of the spins, and the 
SF state corresponds to quasi-long-range ferromagnetic order in the $xy$-plane. 

\section{\label{sec:prot}Protocol: Quantum Crystal Growing}
Starting in the SF regime, we wish to study the adiabatic preparation of the  crystal-like DW insulator state by slowly lowering the tunneling parameter  $J/U$. 
In particular, we are interested in the role played by an inhomogeneity in the form of a parabolic potential $V_\ell$ during this process. 
For this purpose we consider a finite system of $L$ sites described by the Hamiltonian (\ref{eq:Hb}) and characterized by the number of hard-core bosons $N$ and by the scaled trap frequency $\alpha/U$. 
We mimic the finite extent of the simulator region by employing open boundary conditions such that $\ell=-R, -R+1,\ldots,R$ with $R=(L-1)/2$. 
Initially, the tunneling parameter $J/U$ assumes a finite value $(J/U)_0$ and 
the system is prepared in its ground state. 
Then $J/U$ is ramped down to zero at constant rate within a time span of 
duration $T=\tau \hbar/U$. 
In order to quantify the degree of adiabaticity, after this ramp the degree of 
DW order is measured.\footnote{ 
Before evaluating the state and after ramping down the tunneling, one might 
want to add a further step to this protocol in which $\alpha/U$ is ramped down 
to $\alpha/U=0$, such that eventually the system becomes homogeneous for all 
protocols. However, such a step can be omitted; it is irrelevant since at 
$J/U=0$ it will not alter the DW order anymore.}

In order to motivate such a protocol and to gain intuition for the physics related to the presence of the parabolic potential, it is instructive to discuss the protocol described in the preceding paragraph in terms of the local density approximation (LDA).
Introducing the local chemical potential $\mu_\ell=\mu-V_\ell$ one assumes that the ground state of the inhomogeneous system can locally be approximated by the properties of the homogeneous system (summarized in the phase diagram of Fig.~\ref{fig:pd}) with the chemical potential given by $\mu_\ell$.%
\footnote{One condition for the LDA to be valid is that the variation of the trapping potential from site to site should be small compared to the tunneling matrix element $J$, such that particles can delocalize over larger distances (ten sites, say). For $J/U\sim 1$ this leads to the requirement 
$(\alpha/U)R\ll1$. On the other hand, the healing length, the length scale on which a local perturbation influences the many-body wave-function, should be short compared to the spatial structure of the potential.}  
Within the picture of the LDA, the state of an inhomogeneous system with tunneling $J/U$ is represented by a vertical line of finite length (the ``system line'') that cuts through the phase diagram of Fig.~\ref{fig:pd}. 
One end of this line lies at $\mu_{\ell=0}/U=\mu/U$ and corresponds to the 
center of the trap. The other end, to be identified with the edges of the 
system, lies at $\mu_{\ell=\pm R}/U=\mu/U-\alpha R^2/(2U)$. 
So the length of the system line $\Delta\mu/U=(\alpha/U)R^2/2$ is directly 
proportional to $\alpha/U$. 
In the following we will always assume that $\alpha\ge0$ such that the upper end of the system line corresponds to the trap center; results for $\alpha<0$ can be inferred from particle-hole reflection. 

The chemical potential $\mu$ is determined such that the total number of 
particles in the system is given by $N$. 
That is, the system line simply shifts upwards when the particle number 
is increased. When $J/U$ is varied, the chemical potential $\mu$ has to be 
adjusted in order to keep the particle number $N$ fixed. So when we think of 
adiabatically decreasing $J/U$ the system line will move not only leftwards, 
but is displaced also in vertical direction. In Fig.~\ref{fig:pd}(b) this is 
exemplified for three different sets of parameters. The
non-solid lines indicate how $\mu/U$ changes with $J/U$ when the particle 
number is fixed. The vertical lines attached to these lines indicate the 
system line. 

There are good reasons to expect that the presence of a parabolic inhomogeneity might be helpful for the adiabatic preparation of the target state (the DW crystal at $J/U=0$). 
Consider a slow parameter variation following the protocol that is described 
by the short-dashed thin line in Fig.~\ref{fig:pd}(b). When $J/U$ is lowered 
the transition to the DW phase happens first at the center of the trap 
(corresponds to the upper end of the system line that makes contact with the 
DW region first), roughly near $J/U=0.15$. 
From then on, the symmetry-broken DW structure can smoothly grow from the 
center outwards. This process resembles the physics of growing a crystal or 
pulling it out of the melt. However, here, crystallization is not driven 
thermally by lowering the temperature, but rather by quantum fluctuations when 
ramping down the tunneling $J/U$. Hence, one might dub this scheme 
\emph{quantum} crystal growing. 

Growing the DW phase in the inhomogeneous system in this way does not involve 
a sharp phase transition (cf.\ Ref.~\cite{Dziarmaga10} and references 
therein).  Beyond the local density approximation the DW state has a smooth 
boundary in space. When $J/U$ is lowered, this boundary continuously moves 
through the system such that the symmetry broken crystalline order can 
grow\footnote{For simple model systems it was found that a sufficiently slow 
parameter variation guarantees an almost almost adiabatic time evolution in 
such a scenario \cite{DziarmagaRams10a,DziarmagaRams10b}}.

In the presence of the parabolic inhomogeneity the transition is stretched over a finite interval both in the parameter $J/U$ and in time. 
So neither an accurate experimental parameter control nor a precise knowledge of the critical parameter are required to control this process.  
In contrast, for sufficiently large homogeneous system the transition happens  rather suddenly during the ramp when the phase boundary is passed (and it can be expected that the symmetry breaking happens independently in remote places of the system such that defects are created). 

Another advantage of the presence of a parabolic potential is that it allows 
to form a crystal in the center of the trap also for particle numbers below 
half filling. 
The extent $L_\text{DW}$ of the DW crystal will depend on the 
filling $n$ and can be smaller than the extent of the full system, 
$L_\text{DW}=2nL<L$.  In contrast a uniform system away from half filling
does not possess a DW phase. 
  
However, we can also anticipate effects that are not in favor of making the 
system inhomogeneous. For example a finite trap ($\alpha>0$) 
\emph{necessarily} requires filling below $1/2$, which limits the extent of 
the DW crystal to values $L_\text{DW}<L$. There are two different mechanisms 
that lead to such a constraint. The first one is connected to the fact that in 
the superfluid regime the density in the center of the trap will be larger 
than the average density $N/L$. So when $J/U$ is ramped down, for 
$n=1/2$ the DW order will not emerge in the trap center but rather 
independently at those two points (left and right from the center) where the 
local filling is given by $1/2$. As a consequence practically no correlation 
between the crystalline order in both sides of the trap will be established
(a further detrimental effect connected to this scenario will be discussed 
below). Filling factors $N/L$ that avoid this unwanted effect will be lower 
than $1/2$ and such that at $J/U=1/2$ the local density stays below $1/2$ 
everywhere in the trap [this is roughly the case for the protocols with 
$N/L\le0.48$ depicted in Fig.~\ref{fig:pd}(b)].  
 
The second mechanism limiting $L_\text{DW}$ is that the ground state at 
$J/U=0$ at half filling can only be a pure DW crystal of 
length $L_\text{DW}=L$, if the overall potential energy drop within the 
simulator region, $\Delta\mu=\alpha L^2/8$ (the length of the system line), 
stays below $2U$ (the width of the DW lobe at $J/U=0$, see Fig.~\ref{fig:pd}). 
For larger potential drops at the edges and in the center of the trap, regions 
of vacuum or unit filling will form, respectively. In order to avoid a core of 
unit filling in the center of the trap also when $(\alpha/U)L^2>16$, the 
particle number has to be reduced such that that
$L_\text{DW}=2N\le4/\sqrt{\alpha/U}$.

A limitation for achieving an almost adiabatic time evolution in the presence 
of a trap can also be given by mass transport. 
Whereas for the initial state the density decreases smoothly from the center of the trap to the edge, the target state possesses a DW plateau with a filling of 1/2 (i.e.\ with one particle per pair of neighboring sites) in the center for $|\ell|\le L_\text{DW}/2$ and a filling of zero for $|\ell|>L_\text{DW}/2$.
Thus, in order to reach the target state the particle density has to be redistributed.
Therefore a new time scale enters when inhomogeneity is introduced to the system that is not related to the physics of the phase transition, namely the time needed to achieve this redistribution.

This is strikingly evident when the mass flow required in order to achieve the target state is strongly suppressed by the formation of an insulating domain. 
The protocol at half filling (dash-dotted line in Fig.~\ref{fig:pd}) is an example for such a situation. 
When insulating DW domains form and grow at two points in the trap, these domains divide the system into an inner and two outer regions and they become barriers for mass flow between these regions. 
This means that an adiabatic preparation of the target state would require an extremely long ramp time. 
This detrimental mechanism has recently been investigated in the context of the Bose-Hubbard model \cite{BernierEtAl11,NatuEtAl11}. 
Note that also when no insulating barrier appears mass flow can still be a 
factor that determines the time required for the adiabatic preparation of the target state.   
 
In the uniform system ($\alpha/U=0$ and the system line shrinks to a point) at half filling the transition from the SF to the DW phase  happens at the tip of 
the DW lobe and is of second order. 
For a finite harmonic potential, in turn, according to the LDA most parts of 
the system enter the DW phase at a local chemical potential $\mu_\ell\ne0$ and 
therefore at a tunneling parameter $J/U<1/2$ for which the transition is of first order in the (grand canonical) uniform system. 
Of course, corrections to the LDA, as we discussed them already, guarantee that in the presence of the harmonic potential the phase transition is smoothened into a crossover in space (and also in time when the spatial crossover region moves). 
Nevertheless the crossover will be determined by the nature of the phase 
transition it stems from. We can expect that the larger the discontinuity of
the first order transition in the uniform system [quantified by the jump of 
the order parameter $O_\text{DW}$ plotted in Fig.~\ref{fig:pd}(a)] the sharper 
will be the spatial crossover and the smaller will be the rate at which $J/U$ 
can be changed without significantly exciting the system. With respect to this 
effect, steep traps and low filling $N/L$ are not advantageous for an 
adiabatic preparation of the target state. 

\section{\label{sec:sim}Simulation of the time evolution}
In the preceding section we have identified and discussed different mechanisms 
that might play a role when slowly ramping the system from the SF into the DW 
regime in the presence of a parabolic inhomogeneity. While some of them favor 
the presence of the inhomogeneity for the preparation of the 
DW crystal, others disfavor it. In order to find 
out whether (or when) inhomogeneity has a positive or negative influence with 
respect to adiabaticity, we have simulated the protocol described above 
numerically by using the time-dependent matrix 
product state ansatz \cite{Verstraete2008Review, Vidal2004Efficient1D}. 

We consider a realistic system with an odd number of particles $N$ ranging 
from 17 to 31 on $L=61$ sites with open boundary conditions. These odd numbers 
are of course not crucial, but they ensure that the degeneracy between 
different symmetry broken DW patterns is slightly lifted by the parabolic 
potential such that our simulation always leads to a unique reflection 
symmetric pattern with larger site occupation at the even sites 
$\ell=0,\pm2,\pm4,\ldots,\pm(N-1)$. Moreover, also in the absence of the 
parabolic potential an odd number of sites $L$ guarantees a unique 
non-degenerate DW ground state at ``half filling'' $N=(L+1)/2$ (such that the 
DW pattern increases the occupation on both edge sites).

\begin{table}
 \begin{center}
\begin{tabular}{c|ccc}
$\alpha/U$ & Maximum $N$ & $\Delta\mu/U$ & $\alpha R/U$\\
\hline
$1 \times 10^{-4}$ & 31 & 0.045 & $3.0\times 10^{-3}$ \\
$1 \times 10^{-3}$ & 31 & 0.450 & $3.0\times 10^{-2}$  \\
$6 \times 10^{-3}$ & 25 & 2.700 & $1.8\times 10^{-1}$ \\
$1 \times 10^{-2}$ & 19 & 4.500 & $3.0\times 10^{-1}$ 
 \end{tabular}
 \end{center}
 \caption{\label{tab:depths}Summary of potential strengths $\alpha/U$ used in the numerical simulations. Also given the maximum odd particle number $N\le\min\big(2/\sqrt{\alpha/U},31\big)$ that does not lead to unit filling in the trap core at $J/U=0$ for a system of $61$ sites, the potential drop between the center and the edges of the system $\Delta\mu/U$, and the maximum potential difference between neighboring sites $\alpha R/U$. }
\end{table}

In our simulations we compare results for parabolic potentials of four different strengths, $\alpha/U= 10^{-4},10^{-3},6\cdot10^{-3},10^{-2}$.  
We do not switch off the parabolic potential completely, since keeping a small finite potential is required for having a DW phase also away from half filling.
For the two largest values of $\alpha/U$ (the two steepest potentials), we only consider particle numbers $N$ of up to 25 and 19, respectively, that are smaller than $2/\sqrt{\alpha/U}$. This guarantees that the ground state at $J/U=0$ does not possess a core region with unit filling. 
The four different potential strengths $\alpha/U$ give rise to a variation of the local chemical potential from the center to the edge of $\Delta\mu/U=\mu_0/U-\mu_R/U=(\alpha/U)R^2/2 =0.045,0.45,2.7,4.5$, respectively (this is the length of the system line introduced in the preceding section). 
The smallest value is much smaller than the extent of the DW domain in the phase diagram (Fig.~\ref{fig:pd}), the intermediate ones are comparable, and the largest one is much bigger. 
The potential difference between neighboring sites remains smaller than $(\alpha/U)R=3\cdot 10^{-3},3\cdot10^{-2},1.8\cdot10^{-1},3\cdot10^{-1}$, respectively.  Hence, even for the steepest potential at tunneling strength $J/U\ge1/2$ i.e.\ before entering the DW regime, the particles are still delocalized over several sites. The numbers presented in this paragraph are summarized in table \ref{tab:depths}.

The system is initialized in its ground state for $J/U=0.7$, before $J/U$ is 
ramped down linearly from $0.7$ to $0$ within the time span 
$T=\tau \hbar/U$. We choose values between $\tau=1$ (intermediate) and 
$\tau=10$ (moderately large, even larger times would be desirable but are 
numerically costly).
For an Yb168-Yb174 mixture the ramp-time $T$ is thus no larger than $260\ \mathrm{ms}$ using the same estimates as presented in Sec. \ref{sec:sys}.

After the ramp, we compute the distance to the target state of a 
perfect DW crystal with exactly one particle on every other site in the 
central region of the trap. The first measure we consider for this purpose is 
the DW order parameter. However, we are not using $O_\text{DW}$ as it is 
defined in Eq.~(\ref{eq:ODW}), but instead the definition
	\be
	{\tilde O}_\text{DW} = \frac{ \sum_{\ell,\ell'\in M'} \braket{\no_\ell 
	\no_{\ell'}} (-1)^{\ell-\ell'} }{ \left(\sum_{\ell \in M'} 
	\braket{\no_\ell}\right)^2 }
	\ee
which is well defined also for a region $M'$ of finite extent $L'$ only. For a 
perfect zig-zag (DW) structure $3/4$ of the terms in the nominator is $0$ and 
the rest is $1$. The sum in the denominator is $L'/2$ for a perfect zig-zag 
structure so the ratio becomes exactly $1$ in this case. In order to exclude 
edge effects and to be able to compare scenarios with different particle 
numbers, we compute ${\tilde O}_\text{DW}$ based on a region $M'\subset M$ 
containing the central 31 sites ($L'\approx L/2$). In an experiment the density-density correlations
$\braket{\no_\ell \no_{\ell'}}$ entering this parameter can be 
extracted by site resolved measurements \cite{BakrEtAl09,ShersonEtAl10}.

As a second measure we use the nearest-neighbor fidelity $F_\text{NN}$.
We compute the two-site reduced density matrix for each pair of neighboring 
sites $\ell$ and $\ell'$. For the time evolved state $|\psi\ra$ 
this $4\times4$ matrix is defined by
	\be
	\rho^{\ell\ell'}_\psi
	=\trace_{\ell'\in M \setminus \{ \ell,\ell'\} }( \ket\psi\bra\psi ).
	\ee
The two-site reduced density matrix is sufficient to calculate all two-site observables, and therefore characterizes spatially local properties of the state.
We can now calculate the symmetrized overlap 
$d(\rho^{\ell\ell'}_\text{NN},\rho^{\ell\ell'}_\text{DW})$ between the 
two-site density matrices of the time evolved state, $\rho^{\ell\ell'}_\psi$, 
and those computed for the target state with perfect DW order, 
$\rho^{\ell\ell'}_\text{DW}$. The symmetrized overlap between two density 
matrices reads 
\be
 d(\rho_A, \rho_B) = \trace\left( \sqrt{ \sqrt{\rho_A} \rho_B \sqrt{\rho_B} } \right)
\ee
which is $1$ if and only if $\rho_A = \rho_B$ \cite{NielsenChuang}.
The nearest-neighbor fidelity is then defined as the average of these overlaps 
over all neighboring sites in $M'$:
\be
F_\text{NN}\equiv\frac{1}{L'-1} \sum_{\la\ell,\ell'\ra\in M'} d(\rho^{\ell\ell'}_\text{NN},\rho^{\ell\ell'}_\text{DW}).
\ee
 
\begin{figure}
 \centering
   \includegraphics[width=0.8\columnwidth]{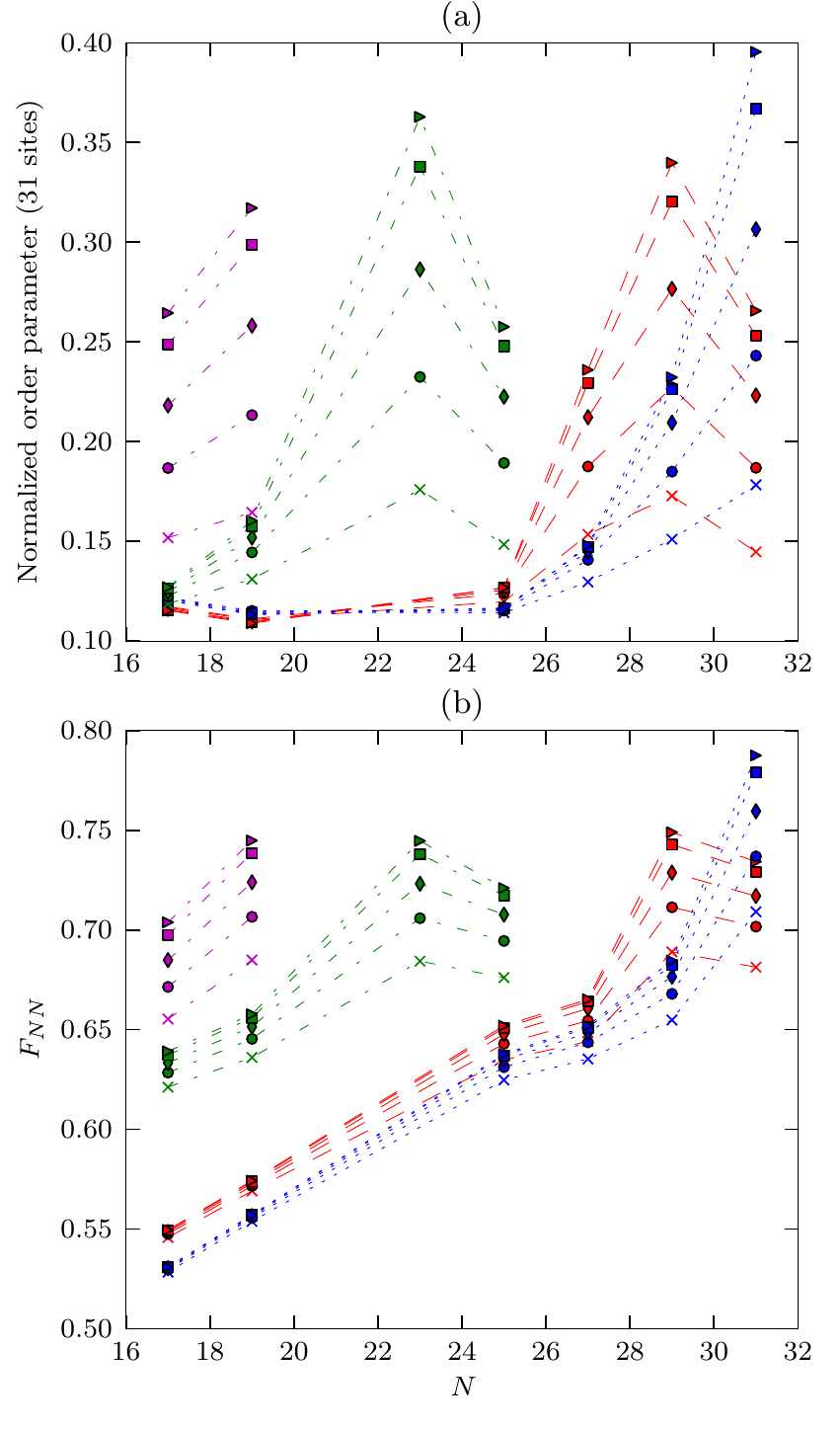}
    \caption{\label{fig:ramps}
Degree of adiabaticity during a ramp from the superfluid to the density 
wave (DW) regime for a system of $N$ particles on a lattice of $L=61$ sites 
in the presence of a parabolic potential of strength 
$\alpha/U= 10^{-4}$ (blue dotted lines), $10^{-3}$ (dashed red lines), 
$6\cdot10^{-3}$ (dash-dotted green lines up to $N=25$), $10^{-2}$ (dash-dotted 
purple lines up to $N=19$).
Starting from the ground state at a tunneling parameter of $J/U=0.7$, the 
time evolution is simulated while $J/U$ is linearly ramped down to zero within 
a time span $\tau\hbar/U$ with $\tau=3$ (crosses), 5 (circles), 7 (diamonds), 
9 (squares), 10 (triangles).
For the final state we plot (a) the normalized DW order parameter 
$\tilde{O}_\text{DW}$ and (b) the nearest-neighbor fidelity 
$F_\text{NN}$ with respect to the DW ground state (b). Both quantities are 
computed for the central region of 31 sites and approach unity in the limit of 
a perfectly adiabatic dynamics. 
The best results are found for the largest particle number $N=31$ 
(corresponding to ``half filling'') in combination with the shallowest 
parabolic potential. 
In contrast, for the lower particle numbers $N\le29$ the presence of steeper 
potentials is always found to be favorable. As 
expected, the degree of adiabaticity increases with $\tau$; the tendency of 
the curves suggest that the results can be improved further by using ramp 
times larger than $\tau=10$.} 
\end{figure}

We plot the main results of our simulation in Fig.~\ref{fig:ramps}. Both 
measures the DW order parameter $\tilde{O}_\text{DW}$ and the nearest-neighbor 
fidelity $F_\text{NN}$ give the same qualitative picture. We find the best 
result (the largest degree of adiabaticity) for the system at ``half 
filling'' with $N=31$ particles in combination with the shallowest 
parabolic potential. However, the degree of adiabaticity that has been 
achieved for generic particle numbers below half filling ($N\le29$) is almost 
as large as for half filling, and it can be increased further by tuning the 
potential depth $\alpha/U$ continuously to its optimal value for every  
particle number (instead of using only four different values of $\alpha/U$ as
we do here). 
So we prefer not to emphasize the better results for half 
filling.
We observe very clearly that the optimal trapping strength 
depends sensitively on the particle number; the optimal depth $\alpha/U$ 
increases when the particle number $N$ is lowered. This effect tells us that a 
finite parabolic inhomogeneity generally does assist the adiabatic preparation 
of the DW crystal. 
As expected, the degree of adiabaticity increases with $\tau$; the tendency of 
the curves suggest that the results can still be considerably improved by 
using ramp times larger than $\tau=10$.

\begin{figure}
\includegraphics[width=1\columnwidth]{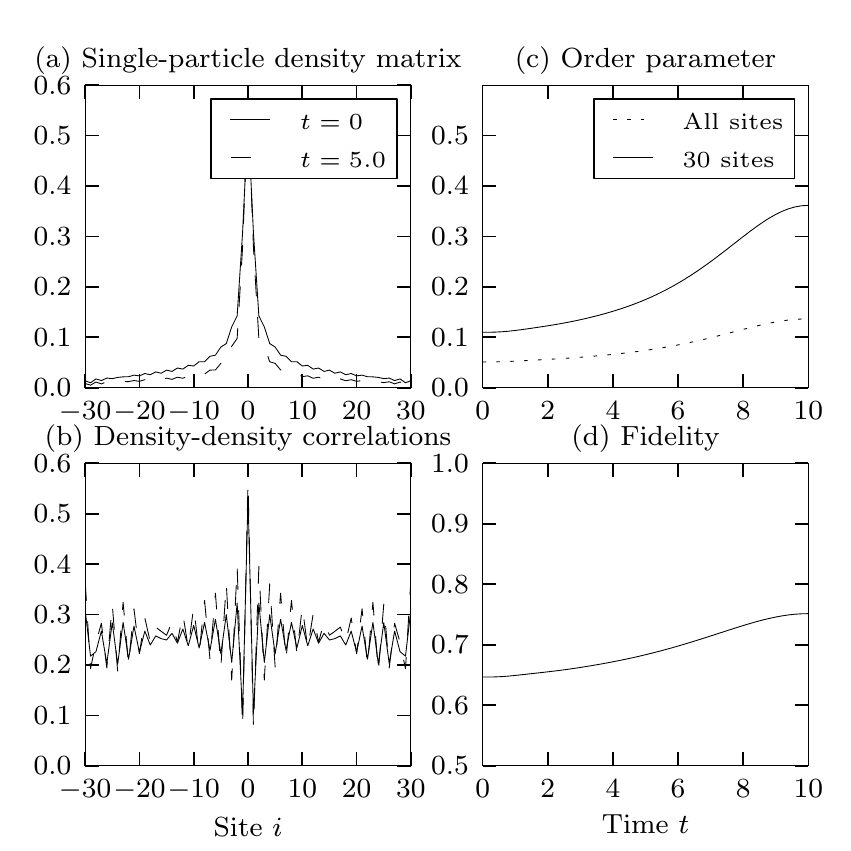}
    \caption{ \label{fig:evolution}
Sample time evolution of a system of $N=29$ particle on $L = 61$ sites
in the presence of the steep parabolic potential of strength 
$\alpha/U = 10^{-3}$. The dimensionless tunneling parameter $J/U$ is linearly 
ramped from $0.7$ to $0$ over a duration of $\tau = 10$ (in units of 
$\hbar/U$). 
(a) Single-particle density matrix $\la\hat{b}^\dag_i\hat{b}_j\ra$ for $j = 0$ 
at times $t = 0$ (solid line) and $t = \tau/2=5$ (dashed line). 
(b) Density-density correlations $\braket{\no_i \no_j}$ for $j = 0$ at
$t = 0$ and $t = \tau/2=5$.
(c) Time evolution of the order parameter $\tilde{O}_\text{DW}$ for the entire 
system and for a 30 site cut-out of the central part of the trap.
(d) Average of the time evolution of the nearest-neighbor density matrix 
fidelity $F_\text{NN}$ for the central part of the trap.
}
\end{figure}

In order to get further insight, in Fig.~\ref{fig:evolution} we report the 
time evolution of the system with $N=29$ and $\alpha/U=10^{-3}$ during the 
ramp with $\tau=10$. Panel (a) and (b) show the single-particle correlations 
$\la\ba_\ell\bo_0\ra$ and the density-density correlations 
$\la\no_\ell\no_0\ra$ both before the ramp (solid lines) and in the middle of 
the ramp (dashed lines). As expected, we can observe that with time the 
single-particle correlations (the off-diagonal 
order) decrease whereas the density-density correlations of the DW type are 
increased. This behavior is also reflected in the fact that the DW order 
parameter $\tilde{O}_\text{DW}$ as well as the nearest-neighbor fidelity 
$F_\text{NN}$ grow with time [panel (c) and (d)].

\begin{figure}   \includegraphics[width=1\columnwidth]{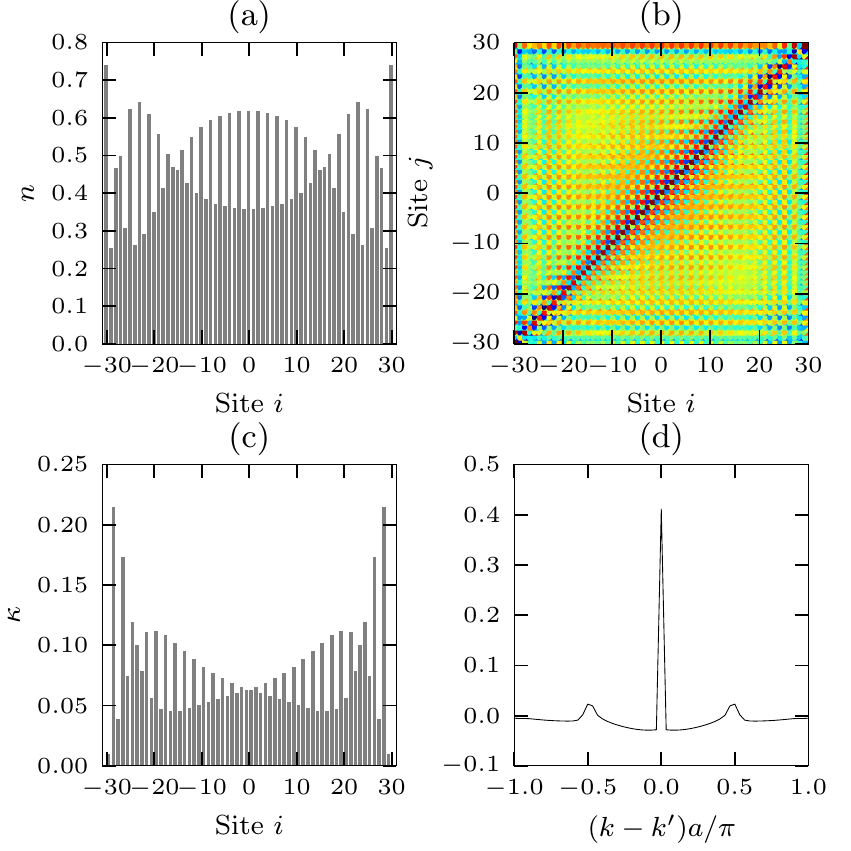}
    \caption{\label{fig:final}
Final state of a system with $N=29$ particles on $L=61$ after a ramp of $J/U$ 
from $0.7$ to 0 with ramp duration $\tau = 10$ and potential strength
$\alpha/U = 10^{-3}$.
(a) Site occupation $\la\no_i\ra$ 
(b) Density-density correlations 
$\braket{n_i n_j}$.
(c) Single-site particle number variance $\kappa_i=\la\no_i^2\ra-\la\no_i\ra^2$
(d) Diagonal momentum correlation function $\braket{a_k^\dagger a_{k'}^\dagger 
a_{k'} a_k}$ as it can be extracted from the shot noise correlations of 
time-of-flight absorption images. The satellite peaks at 0.5 are a signature of 
the density-wave order.}
\end{figure}

\begin{figure*}   \includegraphics[width=1\linewidth]{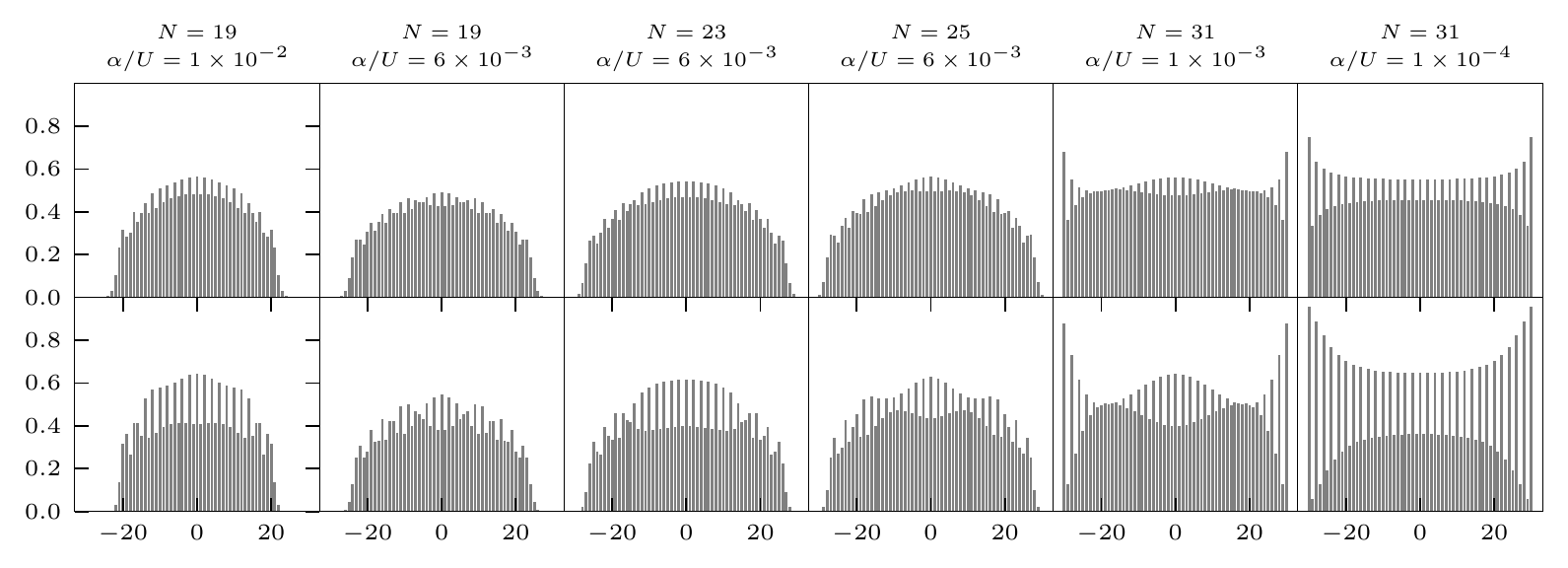} 
    \caption{\label{fig:IniFin}
Density distribution before the ramp (upper row) and after the ramp (lower row) of duration $\tau\hbar/U$ with 
$\tau=10$ for different particle number and trap depths. The low degree of adiabaticity for $N=19$ with $\alpha/U=6\cdot10^{-3}$ can be ascribed to the structure of the initial state. It possesses already weak DW correlations that are contaminated with kinks. The system is not able to get rid of these defects during the ramp. This behavior is found for trap depth (or particle numbers) that are smaller than optimal. The low degree of adiabaticity for both $N=25$ with $\alpha/U=6\cdot10^{-3}$ and $31$ with $\alpha/U=1\cdot10^{-3}$ is related to density modulations on larger scales. This behavior is found for trap depth (or particle numbers) that are larger than optimal.} 
\end{figure*}

A more subtle effect is visible in the density-density correlations 
shown in Fig.~\ref{fig:evolution}(b). Already for the initial state (the 
ground state at $J/U=0.7$, see solid line) it posseses traces of a DW-type 
zig-zag pattern. Superimposed to this pattern one can observe a modulation on 
a larger length scale (comparable to the system size) having nodes roughly at 
$\ell=\pm16$. Having a closer look reveals that at these nodes the zig-zag 
correlations have a kink (where the maxima of the zig-zag pattern shift from even sites to odd sites or vice versa). changes from sites of even index on the one side to sites of odd index on the other side). 
Now, it is very difficult for the system to get rid of the kinks during the 
ramp as it would be required for a perfectly adiabatic evolution (since 
eventually at $J/U=0$ the ground state is a perfect DW). Therefore during the 
ramp the kinks remain, i.e.\ they are converted into defects. This becomes 
evident from Fig.~\ref{fig:final}(a) where the density distribution of the 
time-evolved state after the ramp is plotted (the other subfigures of 
Fig.~\ref{fig:final} display more information on the final state). The 
presence of the kinks explains also the significant drop of the DW order 
parameter $\tilde{O}_\text{DW}$ when computed not only for 31 central sites 
but rather for the whole system [Fig.~\ref{fig:evolution}(c)]. 

For other particle numbers and potential strengths we find similar results. 
Namely the initial ground state at $J/U=0.7$ possesses already a small 
DW-type modulation of the site occupation, typically contaminated with 
superimposed large scale modulations and/or a few kinks. These initial density 
correlations, including the kinks, are amplified when $J/U$ is lowered within 
a time of $\tau\hbar/U$. This behavior can be inferred from 
Fig.~\ref{fig:IniFin} that shows the initial and the final density 
distribution for several particle numbers and trap depths. The best (most 
adiabatic) results are found when there are no kinks or if the kinks lie 
outside the central 31-site region that we use to measure the DW order. 

We find that the results are spoiled by kinks typically when the trapping 
potential is (according to Fig.~\ref{fig:ramps}) shallower than optimal for a 
given particle number (or the particle number is lower than optimal for a 
given potential strength). A typical example for 
kinks spoiling an adiabatic time evolution is found for $N=19$ with 
$\alpha/U=6\cdot 10^{-3}$ (Fig.~\ref{fig:IniFin}).
If, in turn, the trap is too steep for a given particle number (or the particle 
number too large for a given trap depth), we observe superimposed density 
modulations on larger scales in the initial state (not necessarily in 
combination with kinks). These modulations are still found in the time 
evolved state after the ramp, so they lower the degree of adiabaticity. In 
Fig.~\ref{fig:IniFin} this behavior is visible for $N=25$ with 
$\alpha/U=6\cdot 10^{-3}$ and $N=31$ with $\alpha/U=1\cdot10^{-3}$.

The superimposed large-scale modulation of the DW correlations as well as the 
kinks that are present in the initial state and hamper an adiabatic 
time-evolution during the ramp cannot be explained within the simple picture 
of the LDA. They originate from the trap and the 
finite extent of the system. This suggests that the picture that in the 
previous section was drawn on the basis of the LDA (augmented by the 
assumption of smooth crossovers at phase boundaries) does not yet apply 
completely for the experimentally relevant system sizes of 50-100 sites only.

\begin{figure}
  \begin{center}
   \includegraphics[width=\columnwidth]{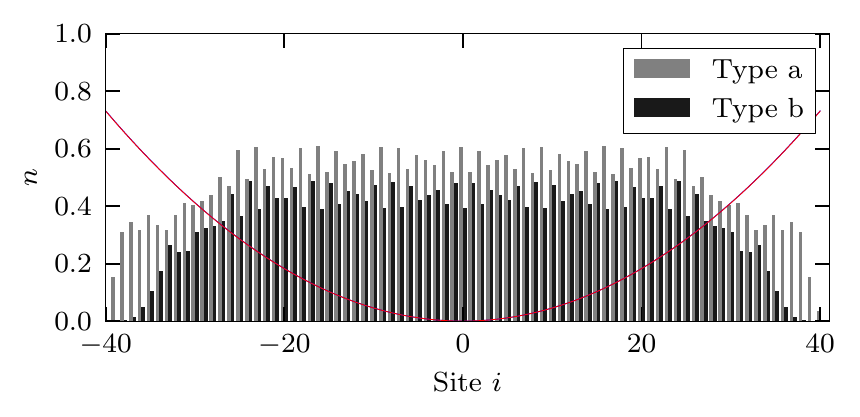} \\
  (a)
   \includegraphics[width=\columnwidth]{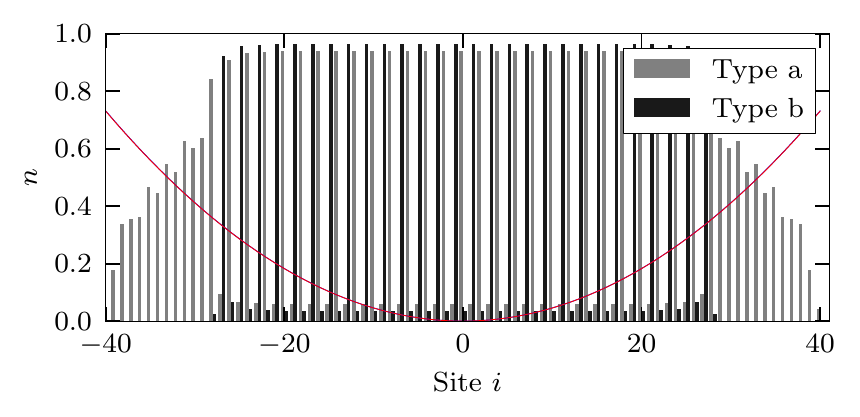} \\
  (b)
  \end{center}
    \caption{\label{fig:full}
    Site atom numbers for the two-species model in the regime used to simulate the XXZ-model with $N_a = 39$ and $N_b = 28$ particles.
The parameters are $U_{ab} = 1.0$, $\alpha_{a/b} \approx 9.1 \times 10^{-4} \pm 6.0 \times 10^{-7}$.
The two-species system has 81 sites and effective tunneling corresponding to 
(a) $J/U = 0.5$ ($J_a = J_b = U_{ab}/8$) and (b) $J/U = 0.12$ ($J_a = U_{ab}/8$, $J_b = J_a/8$). The trap inhomogeneity correspond to 
approximately $\alpha = 10^{-6}$.}
\end{figure}

One might speculate that kinks and density modulations are an artifact of the 
open boundary condition. However, as becomes apparent in 
Fig.~\ref{fig:IniFin}, we also find kinks in the initial states for 
the steep trapping potentials with $\alpha/U=6\cdot10^{-3}$ and 
$\alpha/U=1\cdot10^{-2}$ for which the initial state has practically no 
occupation at the outermost sites such that the boundary conditions do not 
matter. Nevertheless, for the shallower trapping potentials, the initial state 
does depend on the artificial open boundary conditions and the finite size 
effects that we observe here can be modified when the edge of the mixed Mott 
insulator domain $M$ is not approximated by open boundary conditions. A 
more realistic model that captures also the shell surrounding the mixed 
Mott-insulator region $M$ is given by Eq.~(\ref{eq:Heff}). In 
Fig.~\ref{fig:full} we plot the occupation numbers of $a$ and $b$ 
particles for the ground state of this two-species model at $J/U=0.5$ and 
$J/U=0.1$ (the other parameters are specified in the caption). For the small 
tunneling the ground state features defect-free DW/antiferromagnetic order in 
the central Mott region. The small antiferromagnetic correlations found for 
the larger tunneling parameter are, however, contaminated with defects in a 
similar way as observed before when the Mott region $M$ with open boundary 
conditions was treated. These defects can spoil an adiabatic parameter 
variation towards the antiferromagnetic state plotted in 
Fig.~\ref{fig:full}(b).

In an experiment the DW order can be observed either \emph{in situ} using 
high-resolution imaging techniques \cite{BakrEtAl09}, or via time-of-flight 
noise correlation measurements 
\cite{AltmanEtAl04,BachRzazewski04a,FoellingEtAl05} 
probing the two-particle momentum correlation function plotted in 
Fig.~\ref{fig:final}(d). In the latter, the signature of the DW order is given 
by the two satellite peaks. The fact that this feature is rather small is a consequence of the fact that in our simulations the ramp times are still not large enough. 
For larger ramp times (the simulation of which is numerically costly) the 
degree of adiabaticity is still expected to improve considerably. 

\section{\label{sec:con} Conclusion and outlook}

We have pointed out that in a mixed Mott insulator of two bosonic atom species 
in an optical lattice a parabolic inhomogeneity can be created and widely 
tuned (between zero and large finite values) by introducing a finite potential 
difference for both species. And we proposed to use this control knob to 
investigate the role of such an inhomogeneity on the adiabatic preparation of 
an antiferromagnetic state (with a staggered DW pattern for each of the 
species). Numerical simulations of the time evolution of a model describes 
such a system in one dimension lead to the conclusion that a finite 
inhomogeneity generally assists the adiabatic preparation of the bosonic 
antiferromagnet. The optimal strength of the parabolic inhomogeneity depends 
in a sensitive way on the imbalance between the particle numbers of the two 
species; the larger the imbalance (i.e. the smaller the number of hard-core bosons describing the minority species) the larger the optimal strength of the 
inhomogeneity. We find that for a realistic system size (a mixed Mott 
insulator stretched over 60 sites) finite size effects that cannot be 
explained within the local density approximation are significant. Namely the mechanism 
leading to deviations from adiabaticity is related to the presence of 
precursing DW modulations already outside the antiferromagnetic parameter 
regime that -- as a consequence of the finite system size -- comprise 
imperfections like kinks. When the system is ramped into the antiferromagnetic 
regime these modulations, including the imperfections, are amplified. 

We believe that the experimental implementation of tunable parabolic potential 
as we propose it here, can be a valuable tool for finding a good protocol 
for the preparation of antiferromagnetic order. The concept generalizes also 
to two and three spatial dimensions, a situation which can be addressed easily 
in an experiment. A theoretical study of the higher-dimensional case could be 
carried out on a qualitative level using Gutzwiller mean-field theory. 

Another relevant question would be whether such a controllable parabolic 
potential can be useful also for the preparation of antiferromagnetic 
order in a Mott insulator of fermionic atoms 
\cite{SchneiderEtAl08,JoerdensEtAl08}. 

\section*{Acknowledgements}
We acknowledge discussions with Fernando Cuchietti, Massud Haque, and
Philipp Hauke, and thank Maciej Lewenstein for his useful comments on the 
manuscript. SG is grateful to M Lewenstein for his kind hospitality at the 
Institute of Photonic Sciences. Support by ERC AdG QUAGATUA, MINCIN TOQATA, 
FIS2008-00784, and EU IP AQUTE is acknowledged.

\appendix

\section{\label{sec:Bethe}Bethe-Ansatz solution of the homogeneous problem}

The eigenstates of the Hamiltonian (\ref{eq:Hs}) with a homogeneous
magnetic field $h_{\ell} \equiv h$ is found in Refs.\
\cite{YangYang66a,YangYang66b,YangYang66c} using the Bethe Ansatz.

Let $y$ be the total $z$-magnetization density and $\Delta = -U/2J <
0$ characterizes the interaction strength.  Then, $\Delta = -1$
correspond to the Heisenberg anti-ferromagnet as mentioned above.

Instead of finding the groundstate of (\ref{eq:Hs}) for a specific
magnetic field we will instead find the ground state in each total
spin-$z$ subspace, or equivalently, for fixed particle number.  To
find the ground state at a given magnetic field $h$, or chemical
potential $\mu = h$ we should then minimize $u(\Delta, y) - \mu y$,
where $u = E/L$ is the energy density of the ground state, with respect
to $y$, giving $\mu = du/dy(\Delta, y)$.

In \cite{YangYang66c} the phase boundary between the SF and DW phases in Fig. \ref{fig:pd} is given analytically by 
\be
-\frac{\mu}{2J} = 2\sinh(\lambda) \sum_{n=-\infty}^\infty \frac{ (-1)^n }{\cosh(n\lambda)},
\ee
where $\cosh(\lambda) = -\Delta = U/2J$. 

The ground state energy density $u = E/L$ and magnetization $0 < y < 1$ can be found by solving the integral equations \cite[Eq. (7a-c)]{YangYang66b}
\bes
  R(\alpha) &=& \frac{dp}{d\alpha} - \int_{-b}^b K(\alpha, \beta) R(\beta) d\beta \\
  \pi (1 - y) &=& \int_{-b}^b R(\alpha) d\alpha \\
  \frac{u(\Delta, y)}{2J} &=& -\frac{\Delta}{2} - \frac{2}{C} \int_{-b}^b R(\alpha)\frac{dp}{d\alpha} d\alpha
\ees
where $b\in [0, \pi]$ for $\Delta < -1$ and $b\in[0,\infty)$ for $-1 < \Delta < 1$ and the functions $K$ and $dp/d\alpha$ is given in \cite[Table II]{YangYang66b}.

For fixed $b$, the first equation is a Fredholm integral equation of
the second kind, and it can be solved using the Nystrom method
\cite{PressTeukolskyVetterlingFlannery}. For fixed $b$ we can
therefore solve for $R$ and, using this solution, calculate the
magnetization $y$ using the second equation and the energy density $u$
using the third. The magnetic field is $\mu/U = du/dy(\Delta, y)$ which
can also be found by calculating $y$ for $b\pm db$ for a small $db$ numerically and
approximating the derivative using the finite difference.

One can show that the function $R$ is positive, which implies the $y$
is a one-to-one function of $b$. In order to determine the
phase diagram in Fig. \ref{fig:pd} we calculate $y$ and $\mu = de/dy$
as a function of $\Delta$ and $b$. This results in a set of (non-uniformly distributed) 
points $(\Delta, b, y, \mu/U)$, where we
can now plot $y$ as a function of $(\Delta, \mu/U)$, or, $(J/U,
\mu/U)$.

\section*{References}
\bibliography{mybib}

\begin{thebibliography}{10}

\bibitem{Feynman82}
Richard~P. Feynman.
\newblock Simulating physics with computers.
\newblock {\em Int.\ J.\ Theor.\ Phys.}, 21:467, 1982.

\bibitem{HaukeEtAl12}
Philipp Hauke, Fernando~M. Cuchietti, Luca Tagliacozzo, Ivan Deutsch, and
  Maciej Lewenstein.
\newblock Can one trust quantum simulators.
\newblock {\em Rep.\ Prog.\ Phys.}, 75:082401, 2012.

\bibitem{BlochEtAl12}
Immanuel Bloch, Jean Dalibard, and Sylvain Nascimb\`ene.
\newblock Quantum simulations with ultracold quantum gases.
\newblock {\em Nature Phys.}, 8:267, 2012.

\bibitem{Dziarmaga10}
Jacek Dziarmaga.
\newblock Dynamics of a quantum phase transition and relaxation to a steady
  state.
\newblock {\em Adv. Phys.}, 59:1063, 2010.

\bibitem{PolkovnikovEtAl11}
Anatoli Polkovnikov, Krishnendu Sengupta, Alessandro Silva, and Mukund
  Vengalattore.
\newblock Colloquium: Nonequilibrium dynamics of closed interacting quantum
  systems.
\newblock {\em to appear in Rev.\ Mod.\ Phys.\, arXiv:}, 2011.

\bibitem{DziarmagaEtAl99}
Jacek Dziarmaga, Pablo Laguna, and Wojciech~H Zurek.
\newblock Symmetry breaking with a slant: Topological defects after an
  inhomogeneous quench.
\newblock {\em Phys.\ Rev.\ Lett.}, 82:4749, 1999.

\bibitem{DziarmagaRams10a}
Jacek Dziarmaga and Marek~M Rams.
\newblock Dynamics of an inhomogeneous quantum phase transition.
\newblock {\em New J.\ Phys.}, 12:055007, 2010.

\bibitem{DziarmagaRams10b}
Jacek Dziarmaga and Marek~M Rams.
\newblock Adiabatic dynamics of an inhomogeneous quantum phase transition: the
  case of a $z>1$ dynamical exponent.
\newblock {\em New J.\ Phys.}, 12:103002, 2010.

\bibitem{BlochDalibardZwerger08}
Immanuel Bloch, Jean Dalibard, and Wilhelm Zwerger.
\newblock Many-body physics with ultracold gases.
\newblock {\em Rev.\ Mod.\ Phys.}, 80:885, 2008.

\bibitem{LewensteinSanperaAhufinger}
Maciej Lewenstein, Anna Sanpera, and Veronica Ahufinger.
\newblock {\em Ultracold Atoms in Optical Lattices: Simulating quantum
  many-body systems}.
\newblock Oxford University Press, Oxford (UK), 2012.

\bibitem{AltmanEtAl03}
Ehud Altman, Walter Hofstetter, Eugene Demler, and Mikhail~D. Lukin.
\newblock Phase diagram of two-component bosons on an optical lattice.
\newblock {\em N. J. Phys.}, 5:113, 2003.

\bibitem{DuanEtAl03}
L.~M. Duan, E.~Demler, and M.~D. Lukin.
\newblock Controlling spin exchange interactions of ultracold atoms in optical
  lattices.
\newblock {\em Phys.\ Rev.\ Lett.}, 91:090402, 2003.

\bibitem{KuklovSvistunov03}
A.~B. Kuklov and B.~V. Svistunov.
\newblock Counterflow superfluidity of two-species ultracold atoms in a
  commensurate optical lattice.
\newblock {\em Phys.\ Rev.\ Lett.}, 90:100401, 2003.

\bibitem{HubenerEtAl09}
A.~Hubener, M.~Snoek, and W.~Hofstetter.
\newblock Magnetic phases of two-component ultracold bosons in an optical
  lattice.
\newblock {\em Phys.\ Rev.\ B}, 80:245109, 2009.

\bibitem{Powell09}
Stephen Powell.
\newblock Magnetic phases and transitions of the two-species bose-hubbard
  model.
\newblock {\em Phys.\ Rev.\ A}, 79:053614, 2009.

\bibitem{Shrestha10}
Uttam Shrestha.
\newblock Antiferromagnetism in a bosonic mixture of rubidium ($^{87}$rb) and
  potassium ($^{41}$k).
\newblock {\em Phys.\ Rev.\ A}, 82:041603(R), 2010.

\bibitem{CapogrossoSansoneEtAl10}
B.~{Capogrosso-Sansone}, S.~G. S\"oyler, N.~V. Prokof'ev, and B.~Svistunov.
\newblock Critical entropies for magnetic orderingin bosonic mixtures on a
  lattice.
\newblock {\em Phys.\ Rev.\ A}, 81:053622, 2010.

\bibitem{LiEtAl11}
Yongquiang Li, Reza Bakhtiari, Liang He, and Walter Hofstetter.
\newblock Tunable anisotropic magnetism in trapped two-component bose gases.
\newblock {\em Phys.\ Rev.\ B}, 84:144411, 2011.

\bibitem{EckardtEtAl10}
Andr\'e Eckardt, Philipp Hauke, Parvis Soltan-Panahi, Christoph Becker, Klaus
  Sengstock, and Maciej Lewenstein.
\newblock Frustrated quantum antiferromagnetism with ultracold bosons in a
  triangular lattice.
\newblock {\em EPL}, 89:10010, 2010.

\bibitem{SimonEtAl11}
Jonathan Simon, Waseem~S. Bakr, Ruichao Ma, M.~Eric Tai, Philipp~M. Preiss, and
  Markus Greiner.
\newblock Quantum simulation of antiferromagnetic spin chains in an optical
  lattice.
\newblock {\em Nature}, 472:307, 2011.

\bibitem{StruckEtAl11}
J.~Struck, C.~\"Olschl\"ager, R.~{Le Targat}, P.~{Soltan-Panahi}, A.~Eckardt,
  M.~Lewenstein, P.~Windpassinger, and K.~Sengstock.
\newblock Quantum simulation of frustrated classical magnetism in triangular
  optical lattices.
\newblock {\em Science}, 333:996, 2011.

\bibitem{HaukeEtAl12b}
Philipp Hauke, Olivier Tieleman, Alessio Celi, Christoph \"Olschl\"ager,
  Juliette Simonet, Julian Struck, Malte Weinberg, Patrick Windpassinger, Klaus
  Sengstock, Maciej Lewenstein, and André Eckardt.
\newblock Non-abelian gauge fields and topological insulators in shaken optical
  lattices.
\newblock {\em Phys.\ Rev.\ Lett.}, 109:145301, 2012.

\bibitem{LeeEtAl06}
Patrick~A. Lee, Naoto Nagaosa, and Xiao-Gang Wen.
\newblock Doping a mott insulator: Physics of high-temperature superfluidity.
\newblock {\em Rev.\ Mod.\ Phys.}, 78:17, 2006.

\bibitem{HofstetterEtAl02}
W.~Hofstetter, J.~I. Cirac, P.~Zoller, E.~Demler, and M.~D. Lukin.
\newblock High-temperature superfluidity of fermionic atoms in optical
  lattices.
\newblock {\em Phys.\ Rev.\ Lett.}, 89:220407, 2002.

\bibitem{EckardtLewenstein10}
Andr\'e Eckardt and Maciej Lewenstein.
\newblock Controlled hole doping of a mott insulator of ultracold fermionic
  atoms.
\newblock {\em Phys.\ Rev.\ A}, 82:011606(R), 2010.

\bibitem{BernierEtAl11}
Jean-S\'ebastien Bernier, Guillaume Roux, and Corinna Kollath.
\newblock Slow quench dynamics of a one-dimensional bose gas confined to an
  optical lattice.
\newblock {\em Phys.\ Rev.\ Lett.}, 106:200601, 2011.

\bibitem{NatuEtAl11}
Stefan~S. Natu, Kaden R.~A. Hazzard, and Erich~J. Mueller.
\newblock Local versus global equilibration near the bosonic
  mott-insulator–superfluid transition.
\newblock {\em Phys.\ Rev.\ Lett.}, 106:125301, 2011.

\bibitem{HebertEtAl01}
F.~H\'ebert, G.~G. Batrouni, R.~T. Scalettar, G.~Schmid, M.~Troyer, and
  A.~Dorneich.
\newblock Quantum phase transitions in the two-dimensional hardcore boson
  model.
\newblock {\em Phys.\ Rev.\ B}, 65:014513, 2001.

\bibitem{SoerensenEtAl10}
Anders~S. S{\o}rensen, Ehud Altman, Michael Gullans, J.~V. Porto, Mikhail~D.
  Lukin, and Eugene Demler.
\newblock Adiabatic preparation of many-body states in optical lattices.
\newblock {\em Phys.\ Rev.\ A}, 81:061603(R), 2010.

\bibitem{LubaschEtAl11}
Michael Lubasch, Valentin Murg, Ulrich Schneider, J.~Ignacio Cirac, and
  Mari-Carmen Banuls.
\newblock Adiabatic preparation of a heisenberg antiferromagnet using an
  optical superlattice.
\newblock {\em Phys.\ Rev.\ Lett.}, 107:165301, 2011.

\bibitem{Zurek09}
Wojciech~H. Zurek.
\newblock Causality in condensates: Gray solitons as relics of bec formation.
\newblock {\em Phys.\ Rev.\ Lett.}, 102:105702, 2009.

\bibitem{delCampoEtAl10}
A.~{del Campo}, G.~{De Chiara}, G.\ Morigi, M.~B. Plenio, and A.~Retzker.
\newblock Structural defects in ion chains by quenching the external potential:
  The inhomogeneous kibble-zurek mechanism.
\newblock {\em Phys.\ Rev.\ Lett.}, 105:075701, 2010.

\bibitem{delCampoEtAl11}
A.~{del Campo}, A~Retzker, and M.~B. Plenio.
\newblock The inhomogeneous kibble-zurek mechanism: vortex nucleation during
  bose-einstein condensation.
\newblock {\em New J.\ Phys.}, 13:083022, 2011.

\bibitem{HaqueZimmer12}
Masudul Haque and Frank~E. Zimmer.
\newblock Slow interaction ramps in trapped many-particle systems" universal
  deviations from adiabacity.
\newblock {\em arXiv:1110.0840}, 2012.

\bibitem{ColluraKarevski10}
Mario Collura and Dragi Karevski.
\newblock Critical quench dynamics in confined systems.
\newblock {\em Phys.\ Rev.\ Lett.}, 104:200601, 2010.

\bibitem{HuEtAl11}
Anzi Hu, Ludwig Mathey, Eite Tiesinga, Ippei Danshita, Carl~J Williams, and
  Charles~W Clark.
\newblock Detecting paired and counterflow superfluidity via dipole
  oscillations.
\newblock {\em Phys.\ Rev.\ A}, 84:041609(R), 2011.

\bibitem{CataniEtAl08}
J.~Catani, L.~{De Sarlo}, G.~Barontini, F.~Minardi, and M.~Inguscio.
\newblock Degenerate bose-bose mixture in a three-dimensional optical lattice.
\newblock {\em Phys.\ Rev.\ A}, 77:1050, 2008.

\bibitem{MandelEtAl03b}
Olaf Mandel, Markus Greiner, Artur Widera, Tim Rom, Theodor~W. H\"ansch, and
  Immanuel Bloch.
\newblock Controlled collisions for multi-particle entanglement of optically
  trapped atoms.
\newblock {\em Nature}, 425:937, 2003.

\bibitem{LeeEtAl07}
P.~J. Lee, M.~Anderlini, B.~L. Brown, \~J. Sebby-Strabley, W.~D. Phillips, and
  J.V. Porto.
\newblock Sublattice addressing and spin-dependent motion of atoms in a
  double-well lattice.
\newblock {\em Phys.\ Rev.\ Lett.}, 99:020402, 2007.

\bibitem{WeldEtAl09}
David~M. Weld, Patrick Medley, Hirokazu Miyake, David Hucul, David~E.
  Pritchard, and Wolfang Ketterle.
\newblock Spin gradient thermometry for ultracold atoms in optical lattices.
\newblock {\em Phys.\ Rev.\ Lett.}, 103:245301, 2009.

\bibitem{GadwayEtAl10}
Bryce Gadway, Daniel Pertot, Ren\'e Reimann, and Dominik Schneble.
\newblock Superfluidity of interacting bosonic mixtures in optical lattices.
\newblock {\em Phys.\ Rev.\ Lett.}, 105:0031, 2010.

\bibitem{MedleyEtAl11}
Patrick Medley, David~M. Weld, Hirokazu Miyake, David~E. Pritchard, and Wolfang
  Ketterle.
\newblock Spin gradient demagnetization cooling of ultracold atoms.
\newblock {\em Phys.\ Rev.\ Lett.}, 106:195301, 2011.

\bibitem{SoltanPanahiEtAl11}
P.~{Soltan-Panahi}, J.~Struck, P.~Hauke, A.~Bick, W.~Plenkers, G.~Meineke,
  C.~Becker, P.~Windpassinger, M.~Lewenstein, and K.~Sengstock.
\newblock Multi-component quantum gases in spin-dependent hexagonal lattices.
\newblock {\em Nature Phys.}, 7:434, 2011.

\bibitem{FukuharaEtAl09}
Takeshi Fukuhara, Seiji Sugawa, Masahito Sugimoto, Shintaro Taie, and Yoshiro
  Takahashi.
\newblock Mott insulator of ultracold alkaline-earth-metal-like atoms.
\newblock {\em Phys.\ Rev.\ A}, 79:041604(R), 2009.

\bibitem{SugawaEtAl11}
Seiji Sugawa, Kensuke Inaba, Shintaro Taie, Rekishu Yamazaki, Makoto Yamashita,
  and Yoshiro Takahashi.
\newblock Interaction and filling-induced quantum phases of dual mott
  insulators of bosons and fermions.
\newblock {\em Nature Phys.}, 7:642, 2011.

\bibitem{KitagawaEtAl08}
Masaaki Kitagawa, Katsunari Enomoto, Kentaro Kasa, Yoshiro Takahashi, Roman
  Ciury, Pascal Naidon, and Paul~S. Julienne.
\newblock Two-color photoassociation spectroscopy of ytterbium atoms and the
  precise determinations of s-wave scattering lengths.
\newblock {\em Phys.\ Rev.\ A}, 77:012719, 2008.

\bibitem{ThalhammerEtAl09}
G.~Thalhammer, G.~Barontini, J.~Catani, F.~Rabatti, C.~Weber, A.~Simoni,
  F.~Minardi, and M.~Inguscio.
\newblock Collisional and molecular spectroscopy in an ultracold bose–bose
  mixture.
\newblock {\em New J.\ Phys.}, 11:055044, 2009.

\bibitem{Klein73}
D.~J. Klein.
\newblock Degenerate perturbation theory.
\newblock {\em J. Chem. Phys.}, 61:786, 1973.

\bibitem{SpatekOles77}
J.~Spa{\l}ek and A.~M. Ole\'s.
\newblock Ferromagnetism in narrow s-band with inclusion of intersite
  correlations.
\newblock {\em Physica B}, 86-88:375, 1977.

\bibitem{ZouAnderson88}
Z.~Zou and P.~W. Anderson.
\newblock Neutral fermion, charge-e boson excitations in the
  resonating-valence-bond state and superconductivity in la2cuo4-based
  compounds.
\newblock {\em Phys.\ Rev.\ B}, 37:627, 1988.

\bibitem{Boninsegni01}
Massimo Boninsegni.
\newblock Phase separation in mixtrues of hard core bosons.
\newblock {\em Phys.\ Rev.\ Lett.}, 87:087201, 2001.

\bibitem{Boninsegni02}
Massimo Boninsegni.
\newblock Phase separation and stripes in a boson version of a doped quantum
  antiferromagnet.
\newblock {\em Phys.\ Rev.\ B}, 65:134403, 2002.

\bibitem{BoninsengiProkofev08}
Massimo Boninsegni and Nikolay~V. Profkof'ev.
\newblock Phase diagram of an anisotropic bosonic $t$-$j$ model.
\newblock {\em Phys.\ Rev.\ B}, 77:092502, 2008.

\bibitem{LewensteinEtAl04}
M.~Lewenstein, L.~Santos, M.~A. Baranov, and H.~Fehrmann.
\newblock Atomic bose-fermi mixtures in an optical lattice.
\newblock {\em Phys.\ Rev.\ Lett.}, 92:050401, 2004.

\bibitem{YangYang66a}
C.~N. Yang and C.~P. Yang.
\newblock One-dimensional chain of anisotropic spin-spin interactions. i. proof
  of bethe's hypothesis for ground state in a finite system.
\newblock {\em Phys.\ Rev.}, 150:321, 1966.

\bibitem{YangYang66b}
C.~N. Yang and C.~P. Yang.
\newblock One-dimensional chain of anisotropic spin-spin interactions. ii.
  properties of the ground-state energy per lattice site for an infinite
  system.
\newblock {\em Phys.\ Rev.}, 150:327, 1966.

\bibitem{YangYang66c}
C.~N. Yang and C.~P. Yang.
\newblock One-dimensional chain of anisotropic spin-spin interactions. iii.
  applications.
\newblock {\em Phys.\ Rev.}, 151:258, 1966.

\bibitem{Sutherland2004Beautiful}
Bill Sutherland.
\newblock {\em Beautiful Models: 70 Years of Exactly Solved Quantum Many-Body
  Problems}.
\newblock World Scientific, 2004.

\bibitem{Verstraete2008Review}
F.~Verstraete, V.~Murg, and J.~I. Cirac.
\newblock Matrix product states, projected entangled pair states, and
  variational renormalization group methods for quantum spin systems.
\newblock {\em Adv. Phys.}, 57(2):143--224, 2008.

\bibitem{Vidal2004Efficient1D}
Guifre Vidal.
\newblock Efficient simulation of one-dimensional quantum many-body systems.
\newblock {\em Phys.\ Rev.\ Lett.}, 93:040502, 2004.

\bibitem{BakrEtAl09}
W.~S. Bakr, J.~I. Gillen, A.~Peng, S.~Foelling, and M.~Greiner.
\newblock Quantum gas microscope detecting single atoms in a hubbard regime
  optical lattice.
\newblock {\em Nature}, 462:74, 2009.

\bibitem{ShersonEtAl10}
Jacob~F Sherson, Christoph Weitenberg, Manuel Endres, Marc Cheneau, Immanuel
  Bloch, and Stefan Kuhr.
\newblock Single-atom-resolved fluorescence imaging of an atomic mott
  insulator.
\newblock {\em Nature}, 467:68, 2010.

\bibitem{NielsenChuang}
Michael~A Nielsen and Isaac~L. Chuang.
\newblock {\em Quantum Computation and Quantum Information}.
\newblock Cambridge University Press, Cambridge, UK, 2000.

\bibitem{AltmanEtAl04}
Ehud Altman, Eugene Demler, and Mkhail~D. Lukin.
\newblock Probing many-body states of ultracold atoms via noise correlations.
\newblock {\em Phys.\ Rev.\ A}, 70:013603, 2004.

\bibitem{BachRzazewski04a}
Radka Bach and Kazimierz {Rz\c{a}\.zewski}.
\newblock Correlations in atomic systems: Diagnosing coherent superpositions.
\newblock {\em Phys.\ Rev.\ Lett.}, 92:200401, 2004.

\bibitem{FoellingEtAl05}
S.~F\"olling, F.~Gerbier, A.~Widera, O.~Mandel, T.~Gericke, and I.~Bloch.
\newblock Spatial quantum noise interferometry in expanding ultracold atom
  clouds.
\newblock {\em Nature}, 434:481, 2005.

\bibitem{SchneiderEtAl08}
U.~Schneider, L.~Hackerm\"uller, S.~Will, T.~Best, I.~Bloch, T.~A. Costi, R.~W.
  Helmes, D.~Rasch, and A.~Rosch.
\newblock Metallic and insulating phases of repulsively interacting fermions in
  a 3d optical lattice.
\newblock {\em Science}, 322:1520, 2008.

\bibitem{JoerdensEtAl08}
Robert J\"ordens, Niels Strohmaier, Kenneth G\"unter, Henning Moritz, and
  Tilman Esslinger.
\newblock A mott insulator of fermionic atoms in an optical lattice.
\newblock {\em Nature}, 455:204, 2008.

\bibitem{PressTeukolskyVetterlingFlannery}
William~H. Press, Saul~A. Teukolsky, William~T. Vetterling, and Brian~P.
  Flannery.
\newblock {\em Numerical Recipes 3rd Edition: The Art of Scientific Computing}.
\newblock Cambridge University Press, 2007.

\end{thebibliography}
\end{document}